\documentclass[aip,jcp,reprint,noshowkeys,superscriptaddress]{revtex4-2}
\usepackage{graphicx,dcolumn,bm,xcolor,microtype,multirow,amscd,amsmath,amssymb,amsfonts,physics,wrapfig,bbold,siunitx,xspace,braket,txfonts}
\usepackage[version=4]{mhchem}

\usepackage[utf8]{inputenc}
\usepackage[T1]{fontenc}
\usepackage{feynmp-auto}
\usepackage{longtable}
\usepackage{rotating}
\usepackage{hyperref}
\usepackage[compat=1.0.0]{tikz-feynman}

\hypersetup{
    colorlinks,
    linkcolor={red!50!black},
    citecolor={red!70!black},
    urlcolor={red!80!black}
}

\usepackage{listings}
\definecolor{codegreen}{rgb}{0.58,0.4,0.2}
\definecolor{codegray}{rgb}{0.5,0.5,0.5}
\definecolor{codepurple}{rgb}{0.25,0.35,0.55}
\definecolor{codeblue}{rgb}{0.30,0.60,0.8}
\definecolor{backcolour}{rgb}{0.98,0.98,0.98}
\definecolor{mygray}{rgb}{0.5,0.5,0.5}

\definecolor{sqred}{rgb}{0.85,0.1,0.1}
\definecolor{sqgreen}{rgb}{0.25,0.65,0.15}
\definecolor{sqorange}{rgb}{0.90,0.50,0.15}
\definecolor{sqblue}{rgb}{0.10,0.3,0.60}

\lstdefinestyle{mystyle}{
    backgroundcolor=\color{backcolour},
    commentstyle=\color{codegreen},
    keywordstyle=\color{codeblue},
    numberstyle=\tiny\color{codegray},
    stringstyle=\color{codepurple},
    basicstyle=\ttfamily\footnotesize,
    breakatwhitespace=false,
    breaklines=true,
    captionpos=b,
    keepspaces=true,
    numbers=left,
    numbersep=5pt,
    numberstyle=\ttfamily\tiny\color{mygray},
    showspaces=false,
    showstringspaces=false,
    showtabs=false,
    tabsize=2
  }

  \newcolumntype{d}{D{.}{.}{-1}}

\newcommand{\ani}[1]{a_{#1}} 

\newcommand{\fcre}[1]{\Hat{a}_{#1}^{\dag}}
\newcommand{\fani}[1]{\Hat{a}_{#1}}

\newcommand{\bcre}[1]{\Hat{b}_{#1}^{\dag}}
\newcommand{\bani}[1]{\Hat{b}_{#1}}

\newcommand{\bbacre}[1]{\dBar{b}_{#1}^{\dag}}
\newcommand{\bbaani}[1]{\dBar{b}_{#1}}

\newcommand{\bbcre}[1]{\Hat{\beta}_{#1}^{\dag}}
\newcommand{\bbani}[1]{\Hat{\beta}_{#1}}

\usepackage{accents}
\newcommand{\dBar}[1]{\Bar{\Bar{#1}}}

\newcommand{\ERI}[2]{\braket{#1|#2}}

\newcommand{\dbERI}[2]{\mel{#1}{}{#2}}

\newcommand{\Om}{\Omega}

\newcommand{\rev}[1]{\textcolor{black}{#1}}
\newcommand{\revtwo}[1]{\textcolor{black}{#1}}

\newcommand{\fnm}{\footnotemark}
\newcommand{\fnt}{\footnotetext}

\newcommand{\HF}{\text{HF}}
\newcommand{\GW}{\text{$GW$}}

\newcommand{\EOM}{\text{EOM}}

\newcommand{\TCC}{\text{TCC}}
\newcommand{\ECC}{\text{ECC}}

\newcommand{\bb}{\boldsymbol{b}}
\newcommand{\bbb}{\boldsymbol{\beta}}

\newcommand{\bA}{\boldsymbol{A}}
\newcommand{\bB}{\boldsymbol{B}}
\newcommand{\bC}{\boldsymbol{C}}
\newcommand{\bD}{\boldsymbol{D}}

\newcommand{\bX}{\boldsymbol{X}}
\newcommand{\bY}{\boldsymbol{Y}}

\newcommand{\bt}{\boldsymbol{t}}
\newcommand{\bR}{\boldsymbol{R}}
\newcommand{\br}{\boldsymbol{r}}
\newcommand{\bE}{\boldsymbol{E}}
\newcommand{\bM}{\boldsymbol{M}}
\newcommand{\bN}{\boldsymbol{N}}
\newcommand{\bH}{\boldsymbol{H}}
\newcommand{\bO}{\boldsymbol{0}}
\newcommand{\bI}{\boldsymbol{1}}
\newcommand{\bS}{\boldsymbol{S}}
\newcommand{\bz}{\boldsymbol{z}}

\newcommand{\bff}{\boldsymbol{f}}

\newcommand{\bgam}{\boldsymbol{\gamma}}
\newcommand{\bOm}{\boldsymbol{\Omega}}
\newcommand{\bSig}{\boldsymbol{\Sigma}}

\newcommand{\hT}{\Hat{T}}

\newcommand{\hH}{\Hat{H}}
\newcommand{\hbH}{\Bar{H}}
\newcommand{\hbbH}{\dBar{H}}

\newcommand{\hV}{\Hat{V}}
\newcommand{\hZ}{\Hat{Z}}
\newcommand{\hLam}{\Hat{\Lambda}}
\newcommand{\htau}{\Hat{\tau}}

\newcommand{\Ec}{E_\text{c}}

\newcommand{\ii}{\text{i}}
\newcommand{\SupInf}{\textcolor{blue}{Supporting Information}\xspace}

\newcommand{\stkout}[1]{\ifmmode\text{\sout{\ensuremath{#1}}}\else\sout{#1}\fi}

\newcommand{\LCPQ}{Laboratoire de Chimie et Physique Quantiques (UMR 5626), Universit\'e de Toulouse, CNRS, Toulouse, France}
\newcommand{\UnHam}{Department of Chemistry, University of Hamburg, 22761 Hamburg, Germany; The Hamburg Centre for Ultrafast Imaging (CUI), Hamburg 22761, Germany}
\newcommand{\TUW}{Institute for Theoretical Physics, TU Wien, Wiedner Hauptstraße 8-10/136, Vienna, Austria}

\begin{document}	

\title{Connection between $GW$ and Extended Coupled Cluster}

\author{Johannes T\"olle$^{\ast}$}
        \email{johannes.toelle@uni-hamburg.de}
        \affiliation{\UnHam}
	
\author{Marios-Petros \surname{Kitsaras}}
	\affiliation{\LCPQ}

\author{Andreas Irmler}
	\affiliation{\TUW}

\author{Andreas Gr\"uneis}
	\affiliation{\TUW}

\author{Pierre-Fran\c{c}ois \surname{Loos}}
	\affiliation{\LCPQ}
	
\begin{abstract}
\textbf{Abstract:} Coupled-cluster (CC) theory and Green's function many-body perturbation theory (MBPT) have long evolved as distinct yet complementary frameworks for describing electronic correlation.
While CC methods employ exponential wavefunction parametrizations that guarantee size extensivity and systematic improvability, Green's function approaches such as the $GW$ approximation describe quasiparticle and optical excitations through diagrammatic resummations.
Recent analyses have established a formal correspondence between these frameworks: the $GW$ approximation is equivalent to an equation-of-motion (EOM) treatment of the direct-ring coupled-cluster doubles (drCCD) method.
Within this context, the extended CC (ECC) ansatz offers a unified framework connecting CC and MBPT.
This formulation bridges CC-based and Green's function-based methods, providing novel avenues for incorporating vertex corrections within a CC framework that
\rev{preserves a sum-over-state representation of the self-energy and lead to potentially systematically improvable Green's function approaches.}
\bigskip
\begin{center}
	\boxed{\includegraphics[width=0.5\linewidth]{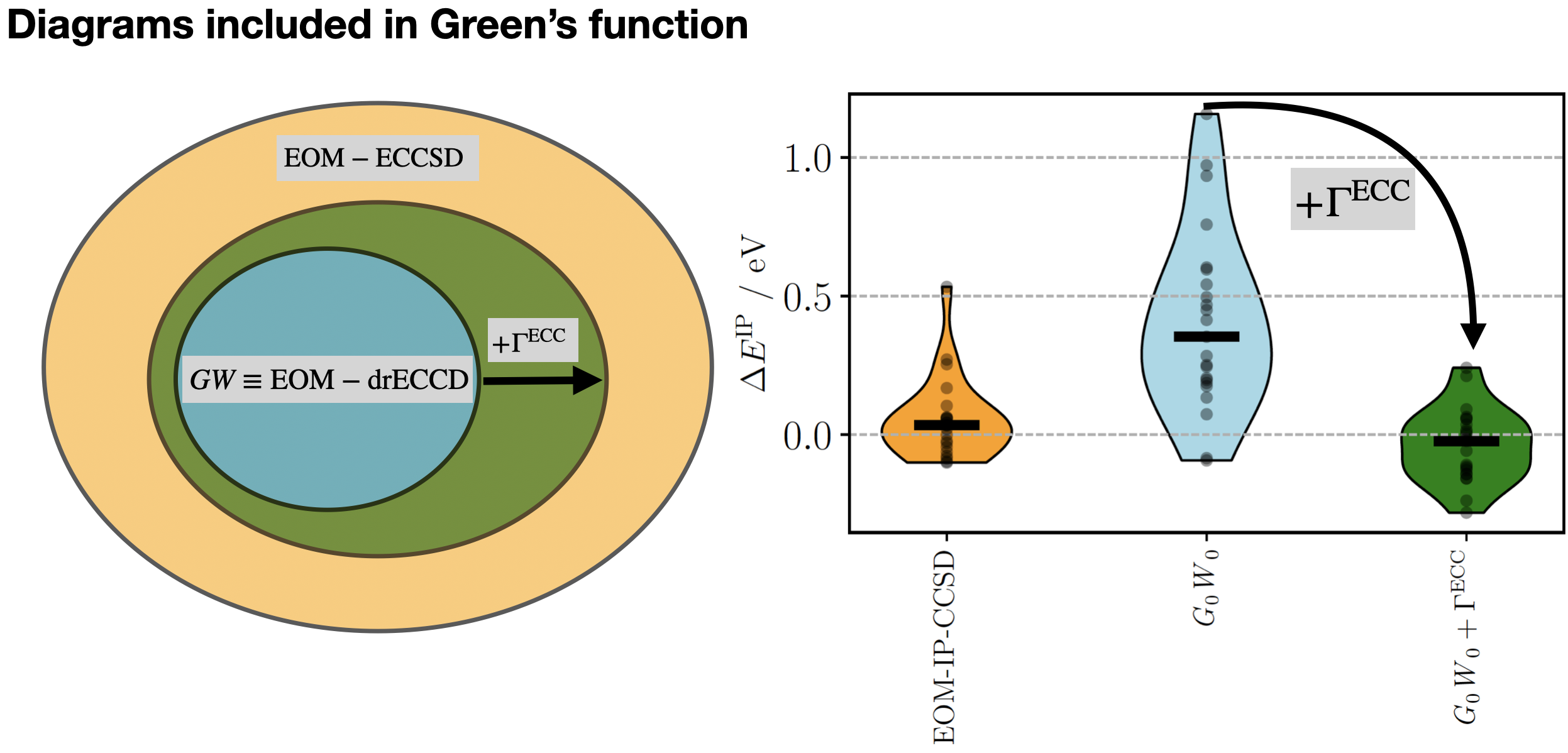}}
\end{center}
\bigskip
\end{abstract}

\maketitle

\section{Introduction}

Coupled-cluster (CC) theory is one of the most mature and powerful methodologies in electronic structure theory.\cite{Crawford_2000,Piecuch_2002,Bartlett_2007,Shavitt_2009,Bartlett_2024} 
Developed over several decades, it now provides access not only to highly accurate ground-state energetics and properties but also to excited states, through either the equation-of-motion (EOM) or linear-response formalisms. \cite{Rowe_1968a,Koch_1990a,Stanton_1993a,Koch_1994,Sneskov_2012} 
Thanks to the sustained efforts of many research groups, CC has become the reference method for high-accuracy electronic structure calculations in molecular systems. \cite{Tajti_2004,Bomble_2006,Harding_2008,Karton_2006,Karton_2017,Goerigk_2011a,Goerigk_2011b,Goerigk_2017,Loos_2020d,Veril_2021,Loos_2025}

Inspired by the seminal work of Hubbard, \cite{Hubbard_1957} the CC ansatz was originally introduced by Coester and Kümmel in the 1950s to describe correlations in nuclear matter. \cite{Coester_1958,Coester_1960} 
It was then brought into quantum chemistry in the 1960s through the seminal work of Sinanoğlu, Čížek, Paldus, and Shavitt. \cite{Sinanoglu_1962,Cizek_1966,Paldus_1972} 
Interestingly, the method later migrated back to nuclear physics in the 1990s \cite{Bishop_1991} and is now widely used for accurate computations of atomic nuclei. \cite{Dean_2004,Kowalski_2004,Hagen_2014} 
More recently, CC has also found applications in condensed matter physics, where methodological and algorithmic advances have enabled its deployment in periodic solids. \cite{Gruber_2018,Zhang_2019a,Wang_2020,Neufel_2023,Masios_2023,Ye_2024,Vo_2024,Moerman_2025a,Moerman_2025b}

Decades of development have revealed both the strengths and limitations of CC theory. 
On the one hand, CC excels at describing ground- and excited-state properties of weakly correlated systems. 
The CCSD(T) model, in particular, is widely regarded as the ``gold standard'' of quantum chemistry, \cite{Purvis_1982,Raghavachari_1989} with systematic extensions such as CCSDT(Q) \cite{Bomble_2005} providing a ``platinum standard'' for even higher accuracy. \cite{Kodrycka_2019}
The introduction of the Lagrangian formalism by Helgaker and co-workers placed the computation of properties as analytic gradients on firm theoretical grounds, opening the way to accurate calculations of both static and frequency-dependent molecular properties of arbitrary order. \cite{Helgaker_1988,Koch_1990b,Koch_1990c,Helgaker_1992,Gauss_2000,Helgakerbook,Hampe_2019} 

On the other hand, the single-reference nature of traditional CC theory becomes problematic in multireference situations where the Hartree--Fock (HF) determinant is not a suitable starting point. 
In such cases, higher-rank excitations (triples, quadruples, etc.) are often required, an EOM-CC approach may be employed, \cite{Musial_2003a,Musial_2003b,Krylov_2008,Shen_2013,Gulania_2019} or one must turn to one of the various flavors of multireference CC. \cite{Musial_2008,Musial_2011,Lyakh_2012,Kohn_2013,Krylov_2017,Evangelista_2018}

A parallel story can be told for Green's function many-body perturbation theory (MBPT). \cite{Onida_2002,Martin_2016} 
Inspired by nuclear physics, with early contributions from Salpeter and Bethe, \cite{Salpeter_1951} Green's function approaches became central in condensed matter physics in the 1960s. 
Hedin's introduction of the $GW$ approximation\cite{Hedin_1965,Aryasetiawan_1998,Reining_2017} was a decisive turning point, as the concept of electronic screening proved remarkably successful in the description of the uniform electron gas. \cite{Lundqvist_1967b,Lundqvist_1967a,Mahan_1989} 
However, $GW$ was not applied to realistic materials until the 1980s. \cite{Strinati_1982a,Strinati_1982b,Hybertsen_1985,Hybertsen_1986,Godby_1986,Godby_1987a}
While Green's function techniques such as the algebraic diagrammatic construction (ADC) \cite{Schirmer_2018,Dreuw_2023} were already being developed for molecules in the 1970s by Cederbaum, Schirmer, von Niessen, Domcke, and coworkers, \cite{Cederbaum_1974,Schirmer_1977,Cederbaum_1977,vonNiessen_1979,Cederbaum_1980,vonNiessen_1981,vonNiessen_1984,Cederbaum_1990} the transfer of the $GW$ and Bethe–Salpeter equation (BSE) formalism to quantum chemistry occurred only more recently. \cite{Shirley_1993,Rohlfing_2000,Stan_2006,Rostgaard_2010,Blase_2011b,Faber_2011,Ke_2011,Bruneval_2012,Bruneval_2013} This was partly because screening effects were thought to be less dominant in small- and medium-sized molecular systems.
Today, however, both $GW$ \cite{Golze_2019,Marie_2024a} and BSE \cite{Blase_2018,Blase_2020} are widely used in molecular science, complementing traditional wavefunction approaches.

In recent years, several groups have begun to explore the theoretical connections between CC theory and Green's function MBPT. 
For example, Lange and Berkelbach analyzed the diagrammatic content of $GW$ and ionization potential (IP) and electron aﬃnity
(EA) EOM-CC theory, \cite{Stanton_1993a,Watts_1994,Musial_2003a,Musial_2003b,Kamiya_2006,Gour_2006} identifying both striking similarities and key differences between the two. \cite{Lange_2018} 
They showed that EOM-CC with singles and doubles (EOM-CCSD) \cite{Purvis_1982,Scuseria_1987,Koch_1990a,Koch_1990c,Stanton_1993a,Stanton_1993b} contains fewer ring diagrams than $GW$, but incorporates a large number of vertex corrections (i.e., beyond ring diagrams) arising from ladder, mixed ring-ladder, and exchange diagrams. 
Including triples yields EOM-CCSDT, \cite{Noga_1987,Scuseria_1988,Watts_1994,Kowalski_2001,Kowalski_2001a,Kucharski_2001}  which includes all diagrams contained in the $GW$ approximation, along with many additional high-order vertex corrections.
Their work builds on earlier insights by Scuseria and co-workers, \cite{Scuseria_2008,Scuseria_2013} who related diagrammatic truncations of CC with doubles (CCD) to variants of the random-phase approximation (RPA) (see also Refs.~\onlinecite{Freeman_1977,Jansen_2010,Peng_2013,Berkelbach_2018,Rishi_2020}).

More recently, Quintero-Monsebaiz \textit{et al.} established connections between BSE@$GW$ and CC at both the ground- and excited-state levels, enabling transfer of methodological insights from one framework to the other. \cite{Quintero_2022} 
They further showed that both $GW$ and BSE can be recast as nonlinear CC-like equations solvable with the standard CC machinery with the same computational scaling. 
In the same spirit, Coveney and Tew have investigated the interconnections between various MBPT schemes and CCD, with a particular focus on the CC self-energy and Green's function.\cite{Coveney_2025a,Coveney_2025b,Coveney_2025c}
Building on the ``upfolded'' version of $GW$ introduced by Bintrim and Berkelbach, \cite{Bintrim_2021} T\"olle and Chan subsequently uncovered an exact connection between $GW$ and the unitary CCD ansatz restricted to the direct-ring (dr) diagrams (drCCD). \cite{Tolle_2022} 
These developments have already borne fruit: One of the authors of the present study derived the first fully analytic $GW$ nuclear gradients, \cite{Tolle_2025} and this work was recently extended by some of us to the first analytic BSE nuclear gradients. \cite{Tolle_2025b} 
Nevertheless, this strategy requires the numerical evaluation of an infinite series of nested (anti)commutators, which is not standard in conventional CC implementations. 
Unfortunately, direct application of EOM-CCD to $GW$ remains hampered by missing correlation effects, as analyzed in Refs.~\onlinecite{Berkelbach_2018,Tolle_2022}.
Very recently, some of us introduced a reformulation of the $GW$ formalism that builds upon the well-established EOM-drCCD framework, \cite{Kitsaras_2026} providing an alternative, fully analytic route to $GW$ nuclear gradients. 
This modified EOM-CCD formulation restores the missing correlation effects inherent to traditional CCD-based approaches while maintaining a consistent and rigorous connection with the $GW$ approximation.

In this context, the extended CC (ECC) ansatz of Arponen offers a promising path forward. \cite{Arponen_1983} 
ECC generalizes standard CC by introducing a bi-variational framework, \cite{Arponen_1982,Arponen_1983,Arponen_1987a,Arponen_1987b} in which both excitation and de-excitation operators are optimized simultaneously (see also Ref.~\onlinecite{Piecuch_1999,Fan_2005,Fan_2006}).
This ansatz preserves size-extensivity while providing a (bi)variational energy functional, making it particularly well-suited for the evaluation of molecular properties. 
In particular, the bi-variational structure ensures that the Hellmann--Feynman theorem can be directly applied, thereby simplifying the computation of response properties. \cite{Pal_1986,Pal_1990,Ghose_1993,Basu-Kumar_1998,Manohar_2004}

Closely related to the ECC framework is the XCC approach, \cite{Bartlett_1988,Bartlett_1989} for which a straightforward EOM variant can be formulated. \cite{Piecuch_1999} Both ECC and XCC employ a doubly similarity-transformed Hamiltonian, which introduces further low-lying correlation contributions, such as important third-order perturbation theory terms that are absent in conventional EOM-CCSD.\cite{Piecuch_1999,Lange_2018}

Comparative benchmark studies of various CC ansätze, including ECC, have been performed by Cooper and Knowles, \cite{Cooper_2010} as well as Evangelista. \cite{Evangelista_2011}
Van Voohris and Head-Gordon introduced the quadratic CCD method via a simplification of the ECCD equations to preserve the computational scaling of the traditional CCD. \cite{vanVoorhis_2000}
In the present contribution, we revisit the relationship between $GW$ and ECC and demonstrate how the latter can provide a natural framework for embedding Green's function formalisms within the CC machinery.

The present work is organized as follows. 
Sections~\ref{sec:GW} and~\ref{sec:GW-EB} introduce the $GW$ approximation using two complementary frameworks: Hedin's approach and the electron-boson formulation, respectively. 
Section~\ref{sec:ECC} briefly reviews the traditional CC energy functional and the extended CC approach, which employs a double similarity transformation that plays a central role in this work. 
Section~\ref{sec:ECC-GW} applies an ECC-like double similarity transformation to the electron-boson Hamiltonian for the $GW$ approximation introduced in Sec.~\ref{sec:GW-EB}.
This yields an ECC electron-boson Hamiltonian, which is represented in Fock space in Sec.~\ref{sec:EOMECC-GW}, yielding an eigenvalue problem for charged excitation energies. 
Section~\ref{sec:EqG0W0} demonstrates the equivalence between the eigenvalue problem of the ECC electron-boson Hamiltonian (Sec.~\ref{sec:EOMECC-GW}) and the $GW$ supermatrix (Sec.~\ref{sec:GW-EB}). 
Section~\ref{sec:Vertexcorrections} identifies the terms in the ECC electron-boson formulation that correspond to vertex corrections beyond $GW$.
In Sec.~\ref{sec:GWLDM}, we motivate a low-order approximation to the $GW$ density matrix within the framework of the double similarity transformed Hamiltonian. 
The resulting linearized one-body density matrix yields an additional correction in the Fock matrix, which can be viewed as an approximation to self-consistent $GW$.
Computational details are summarized in Sec.~\ref{sec:compdet}. 
Numerical results for principal and secondary IPs of a molecular test set are reported in Sec.~\ref{sec:results}, where different levels of theory and vertex corrections are compared. 
Finally, conclusions are drawn in Sec.~\ref{sec:conclusion}.

\section{The $GW$ Approximation}
\label{sec:GW}
The $GW$ approximation is most rigorously formulated within the framework of Hedin's equations, \cite{Hedin_1965} which constitute a self-consistent set of five coupled integro-differential relations connecting the one-body Green's function $G$, the dynamically screened Coulomb interaction $W$, the irreducible vertex function $\Gamma$, the irreducible polarizability $P$, and the exchange-correlation (xc) self-energy $\Sigma_\text{xc}$:
\begin{subequations}
\begin{align}
	\label{eq:Gamma}
	\Gamma(123) & = \delta(12) \delta(13) 
	+  \fdv{\Sigma_\text{xc}(12)}{G(45)} G(46) G(75) \Gamma(673)
	\\
	\label{eq:P}
	P(12) & = - \ii G(13) G(41) \Gamma(342)
	\\
	\label{eq:W}
	W(12) & = v(12) + v(13) P(34) W(42)
	\\
	\label{eq:Sigma_xc}
	\Sigma_\text{xc}(12) & = \ii G(14) W(1^+3) \Gamma(423)
	\\
	\label{eq:G}
	G(12) & = G_\text{H}(12) + G_\text{H}(13) \Sigma_\text{xc}(34) G(42)
\end{align}
\end{subequations}
where $v$ denotes the bare Coulomb interaction, and $G_\text{H}$ is the Hartree (H) Green's function.
Here, integer numbers denote combined space-spin-time variables.

The $GW$ approximation arises upon replacing the three-point irreducible vertex $\Gamma$ by its zeroth-order form, $\Gamma(123) = \delta(12)\delta(13)$, which neglects vertex corrections beyond the independent-particle response. Under this approximation, the irreducible polarizability reduces to
\begin{equation}
	P(12) = -\ii G(12) G(21)
\end{equation}
and the xc self-energy simplifies to
\begin{equation}
	\Sigma_\text{xc}(12) = \ii G(12) W(1^+2)
\end{equation}
\rev{Systematic improvements beyond $GW$ can then, in principle, be obtained by including vertex corrections in either $P$ (internal corrections) and/or $\Sigma_\text{xc}$ (external corrections) in combination with careful diagrammatic analysis to avoid incomplete diagrammatic expansions.} \cite{Shirley_1996,DelSol_1994,Schindlmayr_1998,Morris_2007,Shishkin_2007b,Romaniello_2009a,Romaniello_2012,Gruneis_2014,Hung_2017,Maggio_2017b,Wang_2021a,Mejuto-Zaera_2022,Forster_2022a,Weng_2023,Wen_2024,Bruneval_2024,Forster_2024,Forster_2025}

It is worth noting that both $G$ and $W$ satisfy Dyson-like equations [see Eqs.~\eqref{eq:W} and \eqref{eq:Sigma_xc}, respectively], with $\Sigma_\text{xc}$ and $P$ serving as their respective kernels.
This observation highlights that the $GW$ approximation is inherently a two-step procedure.
In the first step, $P$ is computed to determine the dynamical screening $W$.
In the second step, $\Sigma_\text{xc}$ is evaluated in order to obtain the one-body Green's function $G$.

An alternative and particularly transparent way to view the $GW$ approximation is through its supermatrix representation. \cite{Bintrim_2021,Monino_2022,Quintero_2022,Monino_2023,Tolle_2023,Scott_2023,Marie_2023} 
In this formalism, $GW$ can be written as a linear eigenvalue problem of the form
\begin{equation}
	\bH^\GW \cdot \bR = \bR \cdot \bE
	\label{eq:EigenvaluesG0W0}
\end{equation}
where $\bE$ gathers the charged excitation energies (quasiparticle and satellites), the $GW$ supermatrix reads
\begin{equation} \label{eq:supermatrix}
	\bH^\GW = 
	\begin{pmatrix}
		\bff		&	\bM^{\text{2h1p}}	&	\bM^{\text{2p1h}}
		\\
		\qty(\bM^{\text{2h1p}})^{\dag}	&	\bC^{\text{2h1p}}			&	\bO
		\\
		\qty(\bM^{\text{2p1h}})^{\dag}	&	\bO				&	\bC^{\text{2p1h}}	
	\end{pmatrix}
\end{equation}
and
\begin{equation} \label{eq:GW_eigvec}
	\bR = 
	\begin{pmatrix}
		\br		
		\\
		\br^{\text{2h1p}}
		\\
		\br^{\text{2p1h}}
	\end{pmatrix}
\end{equation}
Here, $\bff$ is the Fock matrix containing the one-hole (1h) and one-particle (1p) configurations, and diagonal sub-blocks defined as 
\begin{align}
	C^\text{2h1p}_{i\nu,i\nu} & = \epsilon_{i} - \Om_{\nu}
	&
	C^\text{2p1h}_{a\nu,a\nu} & = \epsilon_{a} + \Om_{\nu}
\end{align}
for the two-hole-one-particle (2h1p) and two-particle-one-hole (2p1h) configurations, where $\epsilon_{p}$ are the quasiparticle energies.
The corresponding coupling matrices,
\begin{align}
	M^\text{2h1p}_{p,i\nu} & = M_{pi,\nu}
	&
	M^\text{2p1h}_{p,a\nu} & = M_{pa,\nu}
\end{align}
contain the effective two-electron integrals, 
\begin{equation}
\label{eq:sERI_RPA}
	M_{pq,\nu} = \sum_{ia} \ERI{pa}{qi} \qty(\bX + \bY )_{ia,\nu}
\end{equation}
where $\bX$ and $\bY$ are the RPA excitation and deexcitation vectors associated with the one-hole-one-particle (1h1p) RPA (positive) excitation energies $\Omega_\nu$ (see below).
The two-electron integrals $\ERI{pq}{rs}$ are given in Dirac notation, i.e., $\ERI{12}{12}$.
In this work, the indices $p,q,r,s, \dots$ are used for arbitrary orbitals, $i,j,k,l$ label the occupied orbitals, $a,b,c,d$ denote the virtual orbitals, and $\mu, \nu, \dots $ (collective) combined particle-hole indices.
Furthermore, real-valued orbitals are assumed throughout.

The above linear eigenvalue problem can equivalently be recast as a nonlinear quasiparticle equation,
\begin{equation}
	\qty[ \bff + \bSig_\text{c}(\omega) - \omega \bI ] \cdot \br = \bO 
\end{equation}
where the correlation self-energy takes the form
\begin{equation}
\label{eq:self-energy}
\begin{split}
	\bSig_\text{c}(\omega) 
	& = \bM^{\text{2h1p}} \cdot \qty(\omega \bI - \bC^{\text{2h1p}} )^{-1} \cdot \qty(\bM^{\text{2h1p}})^{\dag}
	\\
	& + \bM^{\text{2p1h}} \cdot \qty(\omega \bI - \bC^{\text{2p1h}} )^{-1} \cdot \qty(\bM^{\text{2p1h}})^{\dag} 
\end{split}
\end{equation}
The elements of the correlation part of the self-energy read 
\begin{equation}
	(\Sigma_\text{c})_{pq}(\omega) 
	= \sum_{i\nu} \frac{M_{pi,\nu}M_{qi,\nu}}{\omega - \epsilon_{i} + \Om_{\nu}}
	+ \sum_{a\nu} \frac{M_{pa,\nu}M_{qa,\nu}}{\omega - \epsilon_{a} - \Om_{\nu}}
\end{equation}

The matrices $\bX$ and $\bY$ are obtained by solving the direct RPA eigenvalue problem,
\begin{equation}
	\mqty( \bA & \bB \\ -\bB & -\bA ) \cdot \mqty( \bX & \bY \\ \bY & \bX ) =
	\mqty( \bX & \bY \\ \bY & \bX ) \cdot \mqty( \bOm & \bO \\ \bO & -\bOm )
\end{equation}
where
\begin{subequations} \label{eq:AB_RPA}
	\begin{align}
		A_{ia,jb} &= (\epsilon_a - \epsilon_i) \delta_{ij}\delta_{ab} + \ERI{ib}{aj} 
		\\
		B_{ia,jb} &= \ERI{ij}{ab}
	\end{align}
\end{subequations}
Solving this RPA problem, which scales as $\order*{N^6}$, is a prerequisite for constructing the $GW$ supermatrix.
It can be shown to be equivalent to solving the Riccati equation \cite{Scuseria_2008}
\begin{equation} \label{eq:right_ricatti}
	\bB + \bA \cdot \bt + \bt \cdot \bA + \bt \cdot \bB \cdot \bt = \bO
\end{equation}
with
\begin{equation}
 \bt = \bY \cdot \bX^{-1}
 \label{eq:AmpDef}
\end{equation} 

Although this formal equivalence is useful, it does not by itself provide a clear physical interpretation of the $GW$ approximation.
Nevertheless, it clearly shows that introducing vertex corrections within the same excitation manifold (i.e., up to 2h1p and 2p1h configurations) can only be achieved by modifying one or more of the following quantities in Eq.~\eqref{eq:supermatrix}: the Fock matrix $\bff$, the diagonal blocks $\bC$, or the coupling blocks $\bM$.

\section{$GW$ as a electron-boson problem}
\label{sec:GW-EB}

A complementary and highly insightful perspective on the $GW$ approximation is obtained by recasting it as an electron-boson coupling model. \cite{Lundqvist_1969,Langreth_1970,Hedin_1980,Hedin_1999,Tolle_2023}
\rev{Note that this effectively corresponds to a ``coarse-graining'' of electronic degrees of freedom into quasi-bosonic excitations, a common approximation in the context of boson expansion techniques.\cite{SchuckBook}}
Within this picture, an electron added to or removed from the system interacts linearly with a bath of bosonic excitations representing the collective electronic response. \cite{Tolle_2023}
In this framework, the RPA excitations are interpreted as effective bosonic modes that mediate the dynamical screening of the Coulomb interaction.
The resulting Hamiltonian reads
\begin{equation} \label{eq:HeB}
	\hH_\text{eB} = \hH_\text{e} + \hH_\text{B} + \hV_\text{eB}
\end{equation}
where 
\begin{subequations}
\begin{align}
	\hH_\text{e} 
	& = \sum_{pq} f_{pq} \fcre{p} \fani{q}
	\\
	\label{eq:HB}
	\hH_\text{B}
	& = \sum_{\mu\nu} A_{\mu\nu} \bcre{\mu} \bani{\nu} 
	+ \frac{1}{2} \sum_{\mu\nu} B_{\mu\nu} \qty( \bcre{\mu} \bcre{\nu} + \bani{\mu} \bani{\nu} )
	\\
	\hV_\text{eB} 
	& = \sum_{pq\nu} V_{pq,\nu}  \fcre{p} \fani{q} \qty( \bcre{\nu} + \bani{\nu} )
\end{align}
\end{subequations}
respectively describe the fermionic subsystem, the bosonic bath, and their mutual coupling.
Here, $f_{pq}$ is an element of the Fock matrix, $V_{pq,\nu} \equiv V_{pq,ia} = \ERI{pa}{qi}$ are two-electron integrals that mediate the coupling between the electronic and bosonic degrees of freedom, and the elements of the particle-conserving and particle-non-conserving components of the bosonic Hamiltonian are given by [see Eq.~\eqref{eq:AB_RPA}]
\begin{align}\label{eq:quasibosonindex}
  A_{\mu\nu} & \equiv A_{ia,jb} 
  &
  B_{\mu\nu} & \equiv B_{ia,jb} 
\end{align}

Although neutral electronic excitations are fundamentally composed of fermions, they can often be treated approximately as bosons.
In this framework, quasiboson creation and annihilation operators emulate fermionic particle-hole excitations,
\begin{align}\label{eq:quasiboson}
	\bcre{\nu} & = \fcre{a} \fani{i}
	&
	\bani{\nu} & = \fcre{i} \fani{a}
\end{align}
where $\fcre{a}$ and $\fani{i}$ denote fermionic creation and annihilation operators, respectively.
Because of their underlying fermionic structure, these quasiboson operators do not strictly satisfy fermionic anti-commutation relations anymore, and therefore violate the Pauli exclusion principle and the antisymmetry of the exact electronic wavefunction. \cite{Scuseria_2008}
The quasiboson approximation consists in neglecting these deviations and treating quasibosons as ideal bosons.
This approximation not only simplifies the algebra and the physical interpretation of excitation processes but also transforms otherwise \textit{quartic} fermionic operators, such as the two-body part of the electronic Hamiltonian, into effective (bosonized) \textit{quadratic} Hamiltonians in the quasiboson operators.

Using the composite basis made of the union of the 1h $\{\fani{i}\}$, 1p $\{\fani{a}\}$, 2h1p $\{\fani{i}\bcre{\nu}\}$, and 2p1h $\{\fani{a}\bani{\nu}\}$ configurations, an IP/EA-EOM treatment on the electron-boson Hamiltonian naturally yields the $GW$ supermatrix within the Tamm-Dancoff approximation (TDA). \cite{Bintrim_2021}

To go beyond the TDA, the bosonic Hamiltonian must be diagonalized through a Bogoliubov transformation, which is conveniently expressed as \cite{Tolle_2023}
\begin{equation}
	\hH_\text{B}
	= - \frac{1}{2} \Tr\bA 
	+ \frac{1}{2} \mqty( \bb^\dag & \bb ) \cdot \mqty( \bA & \bB \\ \bB & \bA ) \cdot \mqty( \bb \\ \bb^\dag )
\end{equation}
Introducing the quasiparticle (Bogoliubov) operators \cite{SchuckBook}
\begin{equation}
	\mqty( \bbb \\ \bbb^\dag ) 
	= \mqty( \bX & - \bY \\ - \bY & \bX )^\dag \cdot \mqty( \bb \\ \bb^\dag )
\end{equation}
defined through a unitary transformation built from the RPA eigenvectors, the Hamiltonian becomes
\begin{equation}
	\hH_\text{B} 
	= - \frac{1}{2} \Tr\bA  
	+ \frac{1}{2} \mqty( \bbb^\dag & \bbb ) \cdot \mqty( \bOm & \bO \\ \bO & \bOm ) \cdot \mqty( \bbb \\ \bbb^\dag ).
\end{equation}
The components of the Hamiltonian can then be recast as
\begin{subequations}
\begin{align}
	\hH_\text{B} 
	& = \Ec
	+ \sum_{\mu} \Om_{\mu} \bbcre{\mu} \bbani{\mu} 
	\\
	\hV_\text{eB}
	& = \sum_{\mu\nu} M_{pq,\nu} \fcre{p} \fani{q} \qty( \bbcre{\nu} + \bbani{\nu} )
\end{align}
\end{subequations}
where the RPA correlation energy is
\begin{equation}
	\Ec = \frac{1}{2} \Tr( \bOm - \bA )
\end{equation}
The particle-nonconserving bosonic terms in $\hH_\text{B}$ are thus removed by the Bogoliubov transformation that diagonalizes the RPA problem.
However, full diagonalization is not strictly necessary: a block-diagonalization already suffices to decouple excitation and deexcitation subspaces, rendering the unitary Bogoliubov transformation somewhat excessive.
In the following sections, we will first introduce the extended CC approach and then show that an equivalent block-diagonal structure can be achieved more directly through a double similarity transformation.

\section{Extended Coupled Cluster}
\label{sec:ECC}

The traditional CC (TCC) energy functional is
\begin{equation}
	E_\TCC = \mel{\Phi_0}{(1 + \hLam) e^{-\hT} \hH e^{\hT}}{\Phi_0}
\end{equation}
where $\hT = \sum_q t_q \htau_q$ is an excitation operator written as a function of the general excitation operator $\htau_q$ and $\hLam = \sum_q \lambda_q \htau_q^\dag$ is a deexcitation operator. 
These operators act on the reference wave function $\Phi_0$ to generate excited determinant $\Phi_q$ as $ \ket{\Phi_q} = \htau_q \ket{\Phi_0}$ and $\bra{\Phi_q} = \bra{\Phi_0} \htau_q^\dag$.

The extended CC energy bi-functional is
\begin{equation}
	E_\ECC = \mel{\Phi_0}{e^{\hZ} e^{-\hT} \hH e^{\hT} e^{-\hZ}}{\Phi_0}
\end{equation}
where the linear operator $1 + \hLam$ has been replaced by a proper exponential operator $e^{\hZ}$ with $\hZ = \sum_q z_q \htau_q^\dag$.
In other words, TCC can be seen as an approximation of ECC.
Making the ECC energy functional stationary with respect to the right amplitudes $t_q$ and left amplitudes $z_q$, i.e.,
\begin{align}
	\pdv{E_\ECC}{z_q} & = 0
	& 
	\pdv{E_\ECC}{t_q} & = 0 
	\label{eq:biVar}
\end{align}
yields the amplitude equations 
\begin{subequations}
\begin{align}
	\mel*{\Phi_q}{e^{\hZ} e^{-\hT} \hH e^{\hT}}{\Phi_0} & = 0
	\\
	\mel{\Phi_0}{e^{\hZ} e^{-\hT} \comm{\hH}{\htau_q} e^{\hT}}{\Phi_0} &=0
\end{align}
\end{subequations}
which, contrary to TCC, couple the two sets of amplitudes.
Note that an alternative formulation of the ECC amplitude equations through projection exists, \cite{Piecuch_1999} and numerical results for ground-state energies using both schemes have been shown to be similar. \cite{Fan_2005,Fan_2006}

When restricted to double excitations, that is, $\hT = \hT_2$, the ECC energy functional becomes
\begin{equation}
\begin{split}
	E_\ECC 
	& = \mel{\Phi_0}{e^{\hZ} \qty(\hH e^{\hT_2})_\text{c} }{\Phi_0}
	\\
	& = \mel{\Phi_0}{e^{\hZ} \qty(\hH + \hH \hT_2 + \frac{1}{2} \hH \hT_2^2 + \frac{1}{6} \hH \hT_2^3 + \frac{1}{24} \hH \hT_2^4 )_\text{c} }{\Phi_0}
	\\
	& = \bra{\Phi_0}
	\hH 
	+ \hZ \qty(\hH + \hH \hT_2 + \frac{1}{2} \hH \hT_2^2 )_\text{c}
	\\
	& + \frac{1}{2} \hZ^2 \qty(\frac{1}{2} \hH \hT_2^2 + \frac{1}{6} \hH \hT_2^3)_\text{c}
	+ \frac{1}{6} \hZ^3 \qty(\frac{1}{24} \hH \hT_2^4 )_\text{c}
	 \ket{\Phi_0}
\end{split}
\end{equation}
The subscript c stands for ``connected'' and replaces (nested) commutator(s) between the Hamiltonian and the cluster operator, e.g., $\comm{ \hH }{ \hT_2 } = \qty(\hH \hT_2 )_\text{c}$, as only connected terms survive in these contributions.

At this stage, it is useful to emphasize the similarities and differences between the unitary and extended CC ansätze.
The ECC form, $e^{\hZ} e^{-\hT} \hH e^{\hT} e^{-\hZ}$, can be recast as $e^{-\Hat{t}} \hH e^{\Hat{t}}$ by defining $\Hat{t} = \hT - \hZ + \frac{1}{2}\left[\hZ,\hT\right] + \cdots$, sharing some resemblance with the unitary CC (UCC) expression with $\Hat{t} = \hT - \hT^\dag$.
In UCC, the anti-Hermiticity condition $\Hat{t}^\dag = -\Hat{t}$ ensures a unitary transformation.
In contrast, ECC does not enforce this constraint: $\hZ$ is independent of $\hT$, making $\Hat{t}$ generally non-Hermitian, offering greater flexibility at the cost of Hermiticity. 
Unlike the UCC functional, which results in a non-terminating Baker--Campbell--Hausdorff (BCH) expansion, the ECC approach results in a terminating series, although at a much higher order than the traditional CCD functional.
Lastly, we mention the XCC parametrization that bears the same truncating doubly-similarity transformed Hamiltonian form as ECC but restricts the deexcitations to be the complex conjugate of the excitations $\hZ=\hT^\dagger$. \cite{Piecuch_1999} 
It is argued that due to the decreased flexibility of the XCC ansatz, ECC is expected to recover Hermiticity more effectively in comparison. \cite{Piecuch_1999}

The computational cost of traditional CCD is $\order*{N^6}$, whereas ECCD scales as $\order*{N^{10}}$, which largely explains why it has not been widely adopted despite its attractive formal properties.
Henderson and Scuseria demonstrated that the pair-ECCD method achieves the same $\order*{N^3}$ scaling as the corresponding pair-CCD approach. \cite{Henderson_2014a,Henderson_2015}
Likewise, van Voorhis and Head-Gordon showed that truncating $e^{\hZ}$ after the quadratic term reduces the scaling to $\order*{N^6}$. \cite{vanVoorhis_2000}
These results suggest that suitable restrictions on the double excitation operator can significantly lower the computational cost of ECCD.
Below, we will show that the quadratic nature of the bosonic Hamiltonian underlying $GW$ results in a significant simplification of both the working equations and the associated computational cost.

\section{ECC on the electron-boson Hamiltonian}
\label{sec:ECC-GW}

In this section, we apply the ECC ansatz to the bosonic Hamiltonian $\hH_\text{B}$ defined in Eq.~\eqref{eq:HB}.
As outlined in Sec.~\ref{sec:ECC}, a double similarity transformation is performed:
\begin{equation}
	\hbbH_\text{B} = e^{\hZ} \hbH_\text{B} e^{-\hZ} = e^{\hZ} e^{-\hT} \hH_\text{B} e^{\hT} e^{-\hZ}
\end{equation}
where the bosonic excitation and de-excitation operators are defined as
\begin{align}
	\hT & = \frac{1}{2} \sum_{\mu \nu} t_{\mu \nu} \bcre{\mu} \bcre{\nu}
	&
	\hZ & = \frac{1}{2} \sum_{\mu \nu} z_{\mu \nu} \bani{\mu} \bani{\nu}
\end{align}
Because $\hH_\text{B}$ is quadratic, the BCH expansion associated with the first similarity transformation truncates exactly at second order, yielding
\begin{equation} \label{eq:bH_B}
\begin{split}
	\hbH_\text{B} 
	& = \hH_\text{B} + \comm{ \hH_\text{B} }{ \hT } + \frac{1}{2} \comm{ \comm{ \hH_\text{B} }{ \hT } }{ \hT }
	\\
	& = \frac{1}{2} \sum_{\lambda \sigma} B_{\lambda \sigma} t_{\lambda \sigma} + \sum_{\mu \nu} \bar{A}_{\mu \nu} \bcre{\mu} \bani{\nu} 
	\\ 
	& + \frac{1}{2} \sum_{\mu \nu} \Bar{B}_{\mu \nu} \bcre{\mu} \bcre{\nu} + \frac{1}{2} \sum_{\mu \nu} B_{\mu \nu} \bani{\mu} \bani{\nu} 
\end{split}
\end{equation}
with
\begin{subequations}
\begin{align}
	\Bar{A}_{\mu \nu} & 
	= A_{\mu \nu}
	+ \sum_{\lambda}  t_{\nu \lambda} B_{\lambda \mu}
	\\
	\Bar{B}_{\mu \nu} & 
	= B_{\mu \nu} + \sum_{\lambda} A_{\mu \lambda} t_{\lambda \nu} 
	+ \sum_\lambda t_{\mu \lambda} A_{\lambda \nu} 
	+ \sum_{\lambda \sigma} t_{\mu \lambda } B_{\lambda \sigma} t_{\sigma \nu}
\end{align}
\end{subequations}
where we made use of the fact that $\bt^T = \bt$.\cite{Scuseria_2008}
As seen in Eq.~\eqref{eq:bH_B}, $\hbH_\text{B}$ remains quadratic, containing both number-conserving and non-number-conserving terms.
Hence, the second BCH expansion also terminates at second order, leading to
\begin{equation} \label{eq:bbH_B}
\begin{split}
	\hbbH_\text{B} 
	& = \hbH_\text{B} + \comm{ \hZ }{ \hbH_\text{B} } + \frac{1}{2} \comm{ \hZ }{ \comm{ \hZ }{ \hbH_\text{B} } } 
	\\ 
	& = \frac{1}{2} \sum_{\lambda \sigma} \bar{B}_{\lambda \sigma} z_{\lambda \sigma}
	+ \sum_{\mu \nu} \dBar{A}_{\mu \nu} \bcre{\mu} \bani{\nu} 
	\\
	& + \frac{1}{2} \sum_{\mu \nu} \bar{B}_{\mu \nu} \bcre{\mu} \bcre{\nu} 
	+ \frac{1}{2} \sum_{\mu \nu} \dBar{B}_{\mu \nu} \bani{\mu} \bani{\nu} 
\end{split}
\end{equation}
where
\begin{subequations}
\begin{align}
	\dBar{A}_{\mu \nu} & 
	= \bar{A}_{\mu \nu} 
	+ \sum_{\lambda}  z_{\nu \lambda} \bar{B}_{\lambda \mu}
	\\
	\dBar{B}_{\mu \nu} & 
	= B_{\mu \nu} + \sum_{\lambda} \bar{A}_{\mu \lambda} z_{\lambda \nu} + \sum_\lambda z_{\mu \lambda} \bar{A}_{\lambda \nu} + \sum_{\lambda \sigma} z_{\mu \lambda } \bar{B}_{\lambda \sigma} z_{\sigma \nu}
\end{align}
\end{subequations}

Let $\ket{0_\text{B}}$ denote the bosonic reference vacuum.
The right amplitude equations, obtained as $\bra{0_\text{B}} \bani{\mu} \bani{\nu} \hbbH_\text{B} \ket{0_\text{B}} = 0$, yield the condition $\Bar{B}_{\mu \nu} = 0$, which is equivalent to the (quadratic) Riccati equation in Eq.~\eqref{eq:right_ricatti}.
The left amplitude equations, $\bra{0_\text{B}} \hbbH_\text{B} \bcre{\mu} \bcre{\nu} \ket{0_\text{B}} = 0$, give $\dBar{B}_{\mu \nu} = 0$, which reduces to a set of linear equations in $z_{\mu\nu}$ once $\Bar{B}_{\mu \nu} = 0$:
\begin{equation}
	B_{\mu \nu} + \sum_{\lambda} \bar{A}_{\mu \lambda} z_{\lambda \nu} + \sum_\lambda z_{\mu \lambda} \bar{A}_{\lambda \nu} = 0
\end{equation}
Note that, although left- and right-amplitude equations are generally coupled, they are completely decoupled in the present case. 

Once these two conditions are satisfied, the double similarity-transformed bosonic Hamiltonian simplifies to
\rev{\begin{equation}
		\hbbH_\text{B} = \sum_{\mu \nu} \Bar{A}_{\mu \nu} \bcre{\mu} \bani{\nu} + \frac{1}{2} \sum_{\lambda \sigma} B_{\lambda \sigma} t_{\lambda \sigma} 
\end{equation}}
where all non-number-conserving terms have been eliminated.
\rev{This results in the block-diagonalization of the bosonic Hamiltonian, as discussed in Sec.~\ref{sec:GW-EB}.\cite{Kitsaras_2026}}

Having established the ECC treatment of the bosonic Hamiltonian, we can now turn to the full electron-boson Hamiltonian $\hH_\text{eB}$ given in Eq.~\eqref{eq:HeB}.
The double similarity-transformed electron-boson Hamiltonian reads
\begin{equation}
\begin{split}
	\dBar{H}_\text{eB} 
	& = e^{\hZ} e^{-\hT} \hH_\text{eB} e^{\hT} e^{-\hZ}  
	\\
	& = \hbH_\text{eB} + \comm{ \hZ }{ \hbH_\text{eB} }
	\\ 
	& = \hH_\text{e} + \hbbH_\text{B} + \dBar{V}_\text{eB}
\end{split}
\end{equation}
with 
\begin{equation}
\begin{split}	
	\dBar{V}_\text{eB} 
	& = \sum_{pq} \sum_{\nu} V_{pq,\nu} \fcre{p} \fani{q} \qty( \bcre{\nu} + \bani{\nu} ) \\
	&+ \sum_{pq} \sum_{\nu} \sum_\lambda V_{pq,\lambda} t_{\lambda \nu} \fcre{p} \fani{q} \bcre{\nu} 
	\\
	& + \sum_{pq} \sum_{\nu}  \left[ \sum_\lambda V_{pq,\lambda} z_{\lambda \nu}  + \sum_{\mu \lambda}  V_{pq,\lambda} t_{\lambda \mu} z_{\mu \nu} \right] \fcre{p} \fani{q} \bani{\nu}
\end{split}
\end{equation}

\section{EOM on the ECC electron-boson Hamiltonian}
\label{sec:EOMECC-GW}

Building on the ECC electron-boson Hamiltonian introduced in Sec.~\ref{sec:ECC-GW}, we next derive the EOM formulation for charged excitations.
Within the present ECC-based formalism, the excitation operator $\dBar{R}^{(m)}$ for the $m$th excited state is defined as 
\begin{equation}
\begin{split}
    \dBar{R}^{(m)}
		& = e^{\hat{T}} e^{-\hat{Z}} \hat{R}^{(m)} e^{\hat{Z}} e^{-\hat{T}}\\
		& = \sum_M R^{(m)}_M \dBar{c}_M 
		= \sum_M R^{(m)}_M e^{\hat{T}} e^{-\hat{Z}} \hat{c}_M  e^{\hat{Z}} e^{-\hat{T}} 
		\\ 
        & = \sum_i r^{(m)}_i \ani{i} + \sum_{i\nu} r^{(m)}_{i \nu } \bbacre{\nu} \ani{i}
        + \sum_a r^{(m)}_a \ani{a}  + \sum_{a\nu} r^{(m)}_{\nu a  } \bbaani{\nu} \ani{a}
		\label{eq:ExcOp}
\end{split}
\end{equation}
where $\hat{c}_M$ denotes an excitation operator associated with the $M$th excitation process [see Eq.~\eqref{eq:GW_eigvec}].
The various excitation channels are expressed in terms of double similarity-transformed bosonic creation and annihilation operators,
\begin{subequations}
\begin{align}
	\bbacre{\nu} & = e^{\hT} e^{-\hZ} \bcre{\nu} e^{\hZ} e^{-\hT} 
	\\ 
	\bbaani{\nu} & = e^{\hT} e^{-\hZ} \bani{\nu} e^{\hZ} e^{-\hT}
\end{align}
\end{subequations}
as well as the corresponding left and right reference states in the combined fermionic-bosonic space,
\begin{subequations}
\begin{align}
	\ket{0_\text{e}\dBar{0}_\text{B}} & = e^{\hT} e^{-\hZ} \ket{0_\text{e}0_\text{B}} 
	\\
	\bra{0_\text{e}\dBar{0}_\text{B}} & = \bra{0_\text{e}0_\text{B}} e^{\hat{Z}} e^{-\hat{T}}
\end{align}
\end{subequations}
The use of these transformed creation and annihilation operators is essential to ensure that the EOM formalism satisfies the required ``killer'' conditions. \cite{prasad1985some,datta1993consistent,mukherjee1989effective,Kim_2023,Phillips_2025,Grazioli_2025}

The EOM eigenvalue problem becomes 
\begin{equation}
	\bra{0_\text{e}\dBar{0}_\text{B}} \comm{ \dBar{c}^\dagger_M }{ \comm{ \hH_\text{eB} }{ \dBar{R}^{(m)}} } \ket{0_\text{e}\dBar{0}_\text{B}} 
	= \bra{0_\text{e}\dBar{0}_\text{B}} \comm{ \dBar{c}^\dagger_M }{ \dBar{R}^{(m)} } \ket{0_\text{e}\dBar{0}_\text{B}} E^{(m)}
	\label{eq:DoubleOp}
\end{equation}
with $E^{(m)}$ denoting the energy of the $m$th charged excited state, and $\dBar{c}^\dagger_M$ to the $M$th excitation/deexcitation process [see Eq.~\eqref{eq:ExcOp}], respectively.
Equation \eqref{eq:DoubleOp} can be simplified to 
\begin{equation}
	\bra{0_\text{e}{0}_\text{B}} \comm{ \hat{c}^\dagger_M }{ \comm{ \dBar{H}_\text{eB} }{ \hat{R}^{(m)} } } \ket{0_\text{e}0_\text{B}} 
	= \bra{0_\text{e}{0}_\text{B}} \comm{ \hat{c}^\dagger_M }{ \hat{R}^{(m)} } \ket{0_\text{e}0_\text{B}} E^{(m)}
\end{equation}
and the resulting EOM eigenvalue problem can be written as 
\begin{equation} \label{eq:EOMECCeB}
	\bH^\EOM \cdot \bR = \bR \cdot \bE
\end{equation}
The equivalence with the $G_0W_0$ quasiparticle energies is established analytically in Sec.~\ref{sec:EqG0W0}, identifying the matrix elements of the effective Hamiltonian $\bH^\EOM$ with those of the $GW$ supermatrix $\bH^\GW$ defined in Eq.~\eqref{eq:supermatrix}.

Details regarding the determination of analytic properties within the ECC treatment of the electron-boson Hamiltonian can be found in Ref.~\onlinecite{Kitsaras_2026}.

\section{Equivalence with $G_0W_0$}
\label{sec:EqG0W0}

Having derived the EOM quasiparticle equations using the ECC electron-boson Hamiltonian [see Eq.~\eqref{eq:EOMECCeB}], we now establish their analytical equivalence with the $G_0W_0$ quasiparticle equations.
For this, we rewrite the effective Hamiltonian $\bH^\EOM$ in block form as
\begin{equation} \label{eq:ECCsupermatrix}
	\bH^\EOM = 
	\begin{pmatrix}
		\bff		&	\tilde{\bN}^{\text{2h1p}}	&	\bN^{\text{2p1h}}
		\\
		(\bN^{\text{2h1p}})^\dag	&	\bD^{\text{2h1p}}			&	\bO
		\\
		(\tilde{\bN}^{\text{2p1h}})^\dag	&	\bO				&	\bD^{\text{2p1h}}	
	\end{pmatrix}
\end{equation}
The matrices read 
\begin{subequations}
\begin{align}
	\left[ \tilde{\bN}^{\text{2h1p}} \right]_{pi,\mu} 
	& = \sum_\nu V_{pi,\nu} \qty( \delta_{\nu\mu} + z_{\nu \mu} + \sum_\lambda t_{\nu \lambda} z_{\lambda \mu}  )
	\label{eq:ECC2h1p}
	\\
	\left[ \tilde{\bN}^{\text{2p1h}} \right]_{pa,\mu} 
	& = \sum_\nu V_{pa,\nu} \qty( \delta_{\nu\mu} + z_{\nu \mu} + \sum_\lambda t_{\nu \lambda} z_{\lambda \mu} )
	\label{eq:ECC2p1h}
	\\
	\left[ \bN^{\text{2h1p}} \right]_{pi,\mu} 
	& = \sum_\nu V_{pi,\nu} \qty( \delta_{\nu\mu} + t_{\nu \mu} )
	\\
	\left[ \bN^{\text{2p1h}} \right]_{pa,\mu} 
	& = \sum_\nu V_{pa,\nu} \qty( \delta_{\nu\mu} + t_{\nu \mu} )
\end{align} \label{eq:ECCN}
\end{subequations}
and 
\begin{subequations}\label{eq:DEOMECC}
\begin{align}
	\left[ \bD^{\text{2h1p}} \right]_{i\nu,j\mu}
	& = f_{ij} - A_{\nu \mu} - \sum_{\lambda} t_{\nu \lambda} B_{\lambda \mu}
	\\
	\left[ \bD^{\text{2p1h}} \right]_{a\nu,b\mu} 
	& = f_{ab} + A_{\nu \mu} + \sum_{\lambda} B_{\nu \lambda} t_{\lambda \mu}
\end{align}
\end{subequations}
To demonstrate the equivalence of Eq.~\eqref{eq:EOMECCeB} with the $G_0W_0$ supermatrix [see Eq.~\eqref{eq:supermatrix}], we begin by establishing the following relation $(\bI + \bt) \cdot \bX = \bX + \bY$, which follows directly from Eq.~\eqref{eq:AmpDef}.
Furthermore, we note that $\bI + \bz + \bt \cdot \bz = (\bI - \bt)^{-1}$ (see the \SupInf), simplifying Eqs.~\eqref{eq:ECC2h1p}-\eqref{eq:ECC2p1h}.
From these relations, one finds that $(\bI - \bt)^{-1} \cdot \bX^{-1} = (\bX - \bY)^{-1} = \bX + \bY$, where we made use of $\bt^T=\bt$.\cite{Scuseria_2008}
Finally, one can show that \cite{Kitsaras_2026}
\begin{equation}
	\bX \cdot(\bA + \bt \cdot \bB) \cdot \bX^{-1} = \bOm
\end{equation}

Using the relations derived above, the EOM-ECC eigenvalue problem can be rewritten as 
\begin{equation}
	\bH^\EOM \cdot \bS^{-1} \cdot \bS \cdot \bR = \bR \cdot \bE
\end{equation}
which, after multiplication from the left by the metric  
\begin{equation}
	\rev{
	\bS = \begin{pmatrix}
		\bI & \bO & \bO
		\\
		\bO & \bI^\text{h} \otimes \bX  & \bO
		\\
		\bO & \bO &  \bI^\text{p} \otimes \bX^{-1}
	\end{pmatrix}}
\end{equation}
is identical to the $G_0W_0$ supermatrix formulation of Ref.~\onlinecite{Bintrim_2021,Monino_2022,Quintero_2022,Monino_2023,Tolle_2023,Scott_2023,Marie_2023}, and results in the effective Hamiltonian $\bH^{GW}$ of Eq.~\eqref{eq:supermatrix}.
\\
\rev{Elements of the transformation matrix $\boldsymbol{S}$ are given by
\begin{align}
\left[ \boldsymbol{1}^\text{h} \otimes \boldsymbol{X} \right]_{k\nu,j(ia)} &= \delta_{kj} X_{ia,\nu} 
\end{align}
in the hole-quasiboson block, and
\begin{align}
\left[ \boldsymbol{1}^\text{p} \otimes \boldsymbol{X}^{-1} \right]_{\nu c,(ia) b} &= \delta_{bc} (\boldsymbol{X}^{-1})_{\nu,ia} 
\end{align}
in the particle-quasiboson block.
We have explicitly written down the matrix elements with combined occupied and virtual indices ($ia$) to avoid confusion with collective particle-hole indices $\nu$ arising from transformation in the excitation basis through $\boldsymbol{X}$.}

\section{Beyond-$GW$ vertex corrections from ECC}
\label{sec:Vertexcorrections}

In our opinion, the newly established connection between the ECC formalism and the $GW$ approximation opens exciting avenues for systematically including vertex corrections beyond $GW$.
As noted at the end of Sec.~\ref{sec:GW}, vertex corrections can enter through three distinct components of the $GW$ supermatrix [see Eq.~\eqref{eq:supermatrix}]:
(i) the Fock matrix $\bff$,
(ii) the diagonal blocks $\bC$, and
(iii) the coupling blocks $\bM$.
Reformulating the building blocks $\bC$ and $\bM$ within the ECC framework [see Eq.~\eqref{eq:ECCsupermatrix}] provides direct access to vertex corrections \rev{that preserve the sum-over-state representation of the underlying self-energy. While such a sum-over-state representation is directly linked to causal self-energies for Hermitian effective Hamiltonians, \cite{Bruneval_2025} the non-Hermiticity of the effective Hamiltonian in the context of this work might result in non-causality.
However, similar to the CC Green's function, \cite{Nooijen_1992} we expect such issues to be of minor concern for single-reference states.}
\rev{The sum-over-state representability therefore represents} a notable advantage compared to vertex corrections derived from approximate solutions of Hedin's equations. \cite{Shishkin_2007b,Chen_2015,Ren_2015,Maggio_2017b,Cunningham_2018,Vlcek_2019,Mejuto-Zaera_2022,Rohlfing_2023,Cunningham_2023,Bruneval_2024,Wen_2024,Bruneval_2025,Forster_2025}

In this work, we explore corrections to $\bff$ through static Fock matrix corrections $\bSig(\infty)$ as 
\begin{equation}
    \Sigma_{pq}(\infty) = \sum_{rs} \dbERI{pq}{rs} \qty( \gamma_{pr} - \gamma^\HF_{pr} )
\end{equation}
where $\dbERI{pq}{rs} = \ERI{pq}{rs} - \ERI{pq}{sr}$, $\bgam$ and $\bgam^\HF$ denote the correlated and HF one-body density matrices.
Therefore, the augmented Fock matrix, used in the EOM treatment or, equivalently, in the $GW$ supermatrix, reads 
\begin{equation}
    \bff = \bff^\HF + \bSig(\infty)
\end{equation}
where $\bff^\HF$ is the HF (or alternative mean-field) Fock matrix.
Depending on the choice of $\bgam$, this modification allows for mimicking self-consistency effects \cite{Bruneval_2019_a,Bruneval_2019_b,Bruneval_2021_b} (see Sec.~\ref{sec:GWLDM}), and/or additional vertex corrections beyond $GW$.
\rev{Note that such static self-energy corrections to the density matrix naturally arise in alternative perturbative construction schemes of the self-energy.\cite{vonNiessen_1984}}

Beyond these static corrections, the ECC framework also enables the inclusion of missing exchange contributions in the matrices $\tilde{\bN}$, $\bN$, and $\bD$, which enter $\bH^\EOM$ [see Eq.~\eqref{eq:ECCsupermatrix}]. 
Here, we consider exchange-like contributions that arise from particle-hole contractions of the electron repulsion integrals $\ERI{pq}{rs}$ with the ECC amplitudes ($\bt$ and $\bz$).
The inclusion of such contributions shares similarity with the SOSEX ground-state energy correction to drCCD  proposed in Ref.~\citenum{Gruneis_2009}.

To illustrate how different exchange contributions can be included, we start investigating the diagrammatic representation of the following ring contraction between the electron repulsion integrals $\langle pq | rs\rangle$ with the ECC amplitude $\bt$
\begin{equation}
\begin{split}
    \begin{tikzpicture}
        \begin{feynman}[medium]
            \vertex (a);
            \vertex [right=1.5cm of a] (b) ;
			\vertex [right=1.5cm of b] (testb) ;
            \vertex [below=1.2cm of b] (test) ;
			\vertex [left=1.5cm of test] (testa);
            \vertex [left=0.0cm of test] (c); 
            \vertex [right=1.5cm of c] (d) ;
            \vertex [left=0.5cm of testa] (f1);
            \vertex [right=0.5cm of testa] (f2);
            \vertex [left=0.5cm of testb] (f3);
            \vertex [right=0.5cm of testb] (f4);
            \diagram*  {
            (a) -- [scalar,very thick, edge label=\( \langle ik|ac \rangle \)] (b),
            (b) -- [fermion,half right,very thick, edge label=\( k \)] (c),
            (f1) -- [fermion, very thick,, edge label=\( a \)] (a),
            (a) -- [fermion,very thick, , edge label=\( i \)] (f2),
            (c) -- [fermion,half right,very thick, edge label=\( c \)] (b),
            (c) -- [very thick, edge label'=\( t^{cb}_{kj} \)] (d),
            (f3) -- [fermion,very thick, , edge label'=\( j \)] (d),
            (d) -- [fermion,very thick, edge label'=\( b \)] (f4)
            };
        \end{feynman}
    \end{tikzpicture}\\
	= \sum_{kc} \langle ik|ac \rangle t^{cb}_{kj}
	= \sum_{\lambda} B_{\nu\lambda} t_{\lambda \mu}
\end{split}
\end{equation}
entering the $\bD$ block of $\bH^\EOM$ [see Eqs.~\eqref{eq:ECCsupermatrix} and \eqref{eq:DEOMECC}].
In this case, the particle and hole lines are connected to the same vertices of $\langle ik|ac \rangle$ and $t^{cb}_{kj}$.
Three distinct exchange contributions can be identified by exchanging the vertices to which the particle and hole lines are connected:

(i) Through connection to different vertices of the electron repulsion integral
\begin{equation*}
    \begin{tikzpicture}
        \begin{feynman}[medium]
            \vertex (a);
            \vertex [right=1.5cm of a] (b) ;
			\vertex [right=0.8cm of b] (testb) ;
			\vertex [below=1.5cm of b] (testnew) ;
			\vertex [below=0.5cm of b] (testnew2) ;
            \vertex [left=0.75cm of testnew] (test) ;
			\vertex [left=1.0cm of testnew2] (testa);
            \vertex [left=0.0cm of test] (c); 
            \vertex [right=1.5cm of c] (d) ;
            \vertex [below left=1.1cm of a] (f1);
            \vertex [below left=1.1cm of b] (f2);
            \vertex [left=0.5cm of testb] (f3);
            \vertex [right=0.5cm of testb] (f4);
            \diagram*  {
            (a) -- [scalar,very thick, edge label=\( \langle ik|ca \rangle \)] (b),
            (b) -- [fermion,very thick, edge label=\( k \)] (c),
            (a) -- [fermion, very thick] (f1),
            (f2) -- [fermion,very thick] (b),
            (c) -- [fermion,very thick,, edge label=\( c \)] (a),
            (c) -- [very thick, edge label'=\( t^{cb}_{kj} \)] (d),
            (f3) -- [fermion,very thick] (d),
            (d) -- [fermion,very thick] (f4)
            };
        \end{feynman}
    \end{tikzpicture}
\end{equation*}
(ii) Through connection to different vertices of the amplitude $\bt$
\begin{equation*}
    \begin{tikzpicture}
        \begin{feynman}[medium]
            \vertex (a);
            \vertex [right=1.5cm of a] (b) ;
			\vertex [right=1.5cm of b] (testb) ;
            \vertex [below=1.2cm of b] (test) ;
			\vertex [left=1.5cm of test] (testa);
            \vertex [left=0.0cm of test] (c); 
            \vertex [right=1.5cm of c] (d) ;
            \vertex [left=0.5cm of testa] (f1);
            \vertex [right=0.5cm of testa] (f2);
            \vertex [above left=0.7cm of c] (f3);
            \vertex [right=0.5cm of testb] (f4);
            \diagram*  {
            (a) -- [scalar,very thick, edge label=\( \langle ik|ac \rangle \)] (b),
            (b) -- [fermion,very thick, edge label=\( k \)] (c),
            (f1) -- [fermion, very thick] (a),
            (a) -- [fermion,very thick] (f2),
            (c) -- [fermion,very thick] (f3),
            (c) -- [very thick, edge label'=\( t^{bc}_{kj} \)] (d),
            (d) -- [fermion,very thick, edge label'=\( c \)] (b),
            (f4) -- [fermion,very thick] (d)
            };
        \end{feynman}
    \end{tikzpicture}
\end{equation*}
and (iii) both 
\begin{equation*}
	\begin{tikzpicture}
		\begin{feynman}[medium]
			\vertex (a);
			\vertex [right=1.5cm of a] (b) ;
			\vertex [below=1.5cm of a] (c) ;
			\vertex [below=1.5cm of b] (d) ;
			\vertex [above left=1.5cm of c] (f1);
			\vertex [below left=0.7cm of a] (f2);
			\vertex [above right=1.5cm of d] (f3);
			\vertex [below right=0.7cm of b] (f4);
			\diagram*  {
			(a) -- [scalar,very thick, edge label=\( \langle i c | k a\rangle \)] (b),
			(f1) -- [fermion, very thick] (c),
			(f2) -- [fermion, very thick] (a),
			(c) -- [fermion,very thick, edge label'=\( c \)] (a),
			(b) -- [fermion,very thick, edge label'=\( k \)] (d),
			(c) -- [very thick,very thick, edge label'=\( t^{b c}_{k j} \)] (d),
			(d) -- [fermion, very thick] (f3),
			(b) -- [fermion, very thick] (f4),
			};
		\end{feynman}
	\end{tikzpicture}
\end{equation*}
The latter can be identified as electron-hole-ladder/crossed-ring corrections. \cite{Scuseria_2013,Orlando_2023b}
Additionally, we consider exchange contributions in (i) originating from the contraction with the excitation vector $\hat{R}$ (Sec.~\ref{sec:EOMECC-GW}), i.e., 
\begin{equation}
\begin{split}
	&\sum_{\nu} A_{\nu,\mu} r_{\mu j  } = \sum_{kc} A_{ia,kc} r^{(m)}_{j kc } 
	 \\
	& \rightarrow \sum_{kc} \left( A_{ia,kc}  -  \ERI{ic}{ka} \right) r^{(m)}_{j kc }
\end{split}
\end{equation}
The first term corresponds to the ring contraction
\begin{equation*}
	\begin{tikzpicture}
		\begin{feynman}[medium]
			\vertex (a);
			\vertex [right=1.5cm of a] (b) ;
			\vertex [below=1.5cm of a] (test) ;
			\vertex [above=0.9cm of a] (testb) ;
			\vertex [right=1.5cm of test] (c) ;
			\vertex [right=1.5cm of c] (d) ;
			\vertex [above=0.8cm of c] (f1);
			\vertex [right=0.5 cm of testb] (f2);
			\vertex [above=0.8cm of d] (f3);
			\vertex [left=0.5cm of testb] (f4);
			\diagram*  {
			(a) -- [scalar,very thick, edge label=\( \langle i c | a k\rangle \)] (b),
			(a) -- [fermion, very thick] (f2),
			(c) -- [fermion,very thick,half right, edge label=\( c \)] (b),
			(b) -- [fermion,very thick,half right, edge label=\( k \)] (c),
			(c) -- [very thick,very thick, edge label'=\( r^{(m)}_{j k c} \)] (d),
			(f3) -- [fermion, very thick, edge label=\( j \)] (d),
			(f4) -- [fermion, very thick] (a)
			};
		\end{feynman}
	\end{tikzpicture}
\end{equation*}
and the additional term in the last line represents the contraction of $\ERI{ic}{ka}$ with $r_{j kc }$, where the particle-hole indices ($c$ and $k$) are located at different vertices of the electron repulsion integral:
\begin{equation*}
	\begin{tikzpicture}
		\begin{feynman}[medium]
			\vertex (a);
			\vertex [right=1.5cm of a] (b) ;
			\vertex [below=1.5cm of a] (test) ;
			\vertex [right=0.75cm of test] (c) ;
			\vertex [right=1.5cm of c] (d) ;
			\vertex [above=0.8cm of c] (f1);
			\vertex [above=0.8cm of a] (f2);
			\vertex [above=0.8cm of d] (f3);
			\vertex [above=0.8cm of b] (f4);
			\diagram*  {
			(a) -- [scalar,very thick, edge label=\( \langle i c | k a\rangle \)] (b),
			(a) -- [fermion, very thick] (f2),
			(c) -- [fermion,very thick, edge label=\( c \)] (a),
			(b) -- [fermion,very thick, edge label=\( k \)] (c),
			(c) -- [very thick,very thick, edge label'=\( r^{(m)}_{j k c} \)] (d),
			(f3) -- [fermion, very thick, edge label=\( j \)] (d),
			(f4) -- [fermion, very thick] (b)
			};
		\end{feynman}
	\end{tikzpicture}
\end{equation*}
This example contributes to $\bD$ [see Eq.~\eqref{eq:DEOMECC}].   

To explore the importance of the different exchange corrections, we have implemented the underlying working equations for (i), (ii), and (iii) in spin-adapted form for the EOM effective Hamiltonian of Eq.~\eqref{eq:ECCsupermatrix}.
The resulting equations can be found in the \SupInf.
In the following, we will denote the different exchange corrections as $\Gamma_\text{V}^\text{x}$ for case (i), $\Gamma_\text{A}^\text{x}$ for case (ii), and $\Gamma_\text{CR}^\text{x}$ for case (iii).
Combinations of these corrections will be indicated by the addition of the respective labels, e.g., $\Gamma_\text{V+A}^\text{x}$ indicates the inclusion of both (i) and (ii) contributions.

We would like to stress that the proposed corrections (i) to (iii) are a choice made in the present work.
The corrections are motivated also by their potential for low-scaling implementations. 
While the computational cost of the present approach is $\order*{N^6}$, we would like to emphasize that the cost for determining the amplitudes can be reduced to $\order*{N^4}$ through techniques such as Cholesky decomposition. \cite{Scuseria_2008,Tolle_2023}
Furthermore, in combination with the resolution-of-the-identity (RI) technique, \cite{Beebe_1977,Pedersen_2024} the matrix-vector products for the EOM eigenvalue problem, including exchange corrections, can be reduced to $\order*{N^5}$, whereas application of tensor-hypercontraction techniques (THC) \cite{hohenstein2012tensor,Parrish_2012} reduces this scaling further to $\order*{N^4}$.
The applicability of these low-scaling techniques within the vertex corrections presented here will be explored in future work.

Beyond that, the choice made here can also be rationalized by comparison to alternative $GW$ vertex corrections proposed in the literature (see Refs.~\onlinecite{Shishkin_2007b,Chen_2015,Ren_2015,Maggio_2017b,Cunningham_2018,Vlcek_2019,Mejuto-Zaera_2022,Rohlfing_2023,Cunningham_2023,Bruneval_2024,Wen_2024,Bruneval_2025,Forster_2025}).
Within Hedin's equations [see Eqs.~\eqref{eq:Gamma}-\eqref{eq:G}], vertex correction enter both the self-energy $\Sigma$ [see Eq.~\eqref{eq:Sigma_xc}] and the polarizability $P$ [see Eq.~\eqref{eq:P}].
Note that the vertex itself is defined through the functional derivative of $\Sigma$ with respect to $G$ [see Eq.~\eqref{eq:Gamma}].
This formal structure has motivated a variety of practical schemes, commonly classified according to whether vertex corrections are included in $\Sigma$, in $P$, or in both. 
Moreover, different approximations to the vertex $\Gamma$ have been explored, \cite{Maggio_2017b,Vlcek_2019,Rohlfing_2023,Wen_2024,Forster_2025} many of which suffer from \rev{a lack of causality.} \cite{Pavlyukh_2016,Bruneval_2024,Bruneval_2025}

The choices made in the present work can be interpreted as modifying both $\Sigma$ and $P$ such that the resulting (non-Hermitian) self-energy allows for a sum-over-state representation.
However, several alternative, potentially systematically improvable, vertex corrections can be devised within the ECC framework, which we leave for future work.
For example, of particular interest are the inclusion of additional contributions in the amplitude equations themselves and the effect of particle-particle and hole-hole ladder contributions at the EOM level.

\section{$GW$ linearized density matrix from ECC perturbation theory}
\label{sec:GWLDM}

As noted in Sec.~\ref{sec:Vertexcorrections}, the inclusion of static Fock matrix corrections $\bSig(\infty)$ allows for mimicking self-consistency effects within the $GW$ approximation.
Here, we derive the $GW$ linearized one-body density matrix $\bgam$ within the ECC framework, which can be used to compute $\bSig(\infty)$.

References \onlinecite{Bruneval_2019_a,Bruneval_2019_b,Bruneval_2021_b} derived the linearized one-body density matrix within the $GW$ approximation as
\begin{subequations}
\begin{align}
	\gamma_{ij} & = \delta_{ij} - \sum_{a\nu}\frac{M_{ia,\nu} M_{ja,\nu}}{\qty(\epsilon_i - \epsilon_a - \Omega_\nu )\qty(\epsilon_j - \epsilon_a - \Omega_\nu )} 
	\label{eq:OccOccBlockD}
	\\
	\gamma_{ab} & = \sum_{i\nu}\frac{M_{ai,\nu} M_{bi,\nu}}{\qty(\epsilon_i - \epsilon_a - \Omega_\nu )\qty(\epsilon_i - \epsilon_b - \Omega_\nu )} 		
	\label{eq:VirtVirtBlockD}
	\\
	\gamma_{ia} & = \frac{f_{ia}}{\epsilon_i - \epsilon_a} \nonumber \\
	&+ \frac{1}{\epsilon_i - \epsilon_a} \qty[ \sum_{b\nu} \frac{M_{ib,\nu} M_{ab,\nu}}{\epsilon_i - \epsilon_b - \Omega_\nu} 
	+ \sum_{j\nu} \frac{M_{ij,\nu} M_{aj,\nu}}{\epsilon_j - \epsilon_a - \Omega_\nu} ]
	\label{eq:OccVirtBlockD}
\end{align}
\end{subequations}
In the following, we show how these equations can be derived within the ECC framework perturbatively.
For this, we partition the Hamiltonian as 
\begin{equation}
	\hH(\lambda) = \hH_0 + \lambda \hV
\end{equation}
where $\lambda$ denotes the perturbation parameter, and
\begin{subequations}
\begin{align}
	\hH_0 & = \sum_p f_{pp} \fcre{p} \fani{p} + \sum_{\mu \nu} \dBar{A}_{\mu \nu} \bcre{\nu} \bani{\mu}
	\\
	\hV & = \sum_{pq\nu} \dBar{M}_{pq\nu} \fcre{p} \fani{q} \qty( \bcre{\nu} + \bani{\nu} ) + \sum_{p \neq q} f_{pq} \fcre{p} \fani{q}
\end{align}
\end{subequations}
Within perturbation theory, \cite{Kutzelnigg_2009} the doubly similarity-transformed Hamiltonian reads
\begin{equation}
	\dBar{H}(\lambda) = e^{\hZ (\lambda )} e^{-\hT(\lambda)} \hH(\lambda) e^{\hT(\lambda)} e^{-\hZ(\lambda)}
\end{equation}
where the excitation and deexcitation operators read
\begin{subequations}
\begin{align}
	\hT(\lambda) & = \lambda \hT^{(1)} + \lambda^2 \hT^{(2)} + \cdots
	\\
	\hZ(\lambda) & = \lambda \hZ^{(1)} + \lambda^2 \hZ^{(2)} + \cdots
\end{align}
\end{subequations}
with
\begin{subequations}
\begin{align}
	\hT^{(n)} &= \hT^{(n)}_1 + \hT^{(n)}_2 =  \sum_{ia} t^{(n)}_{ia} \fcre{a} \fani{i} + \sum_{ia\nu} t^{(n)}_{ia\nu} \fcre{a} \fani{i} \bcre{\nu}
	\\
	\hZ^{(n)} &= \hZ^{(n)}_1 + \hZ^{(n)}_2 = \sum_{ia} z^{(n)}_{ia} \fcre{i} \fani{a} + \sum_{ia\nu} z^{(n)}_{ia\nu} \bani{\nu} \fcre{i} \fani{a}
\end{align}
\end{subequations}
and the $n$th-order amplitude equations can be deduced from  
\begin{subequations}
\begin{align}
	\hT^{(n)}_k  \rightarrow \frac{1}{n!} \eval{ \pdv[n]{\bra{k} \dBar{H}(\lambda) \ket{0}} {\lambda} }_{\lambda=0} = 0
	\\
	\hZ^{(n)}_k  \rightarrow  \frac{1}{n!} \eval{ \pdv[n]{\bra{0} \dBar{H}(\lambda) \ket{k}} {\lambda} }_{\lambda=0} = 0
\end{align}
\end{subequations}
where $k \in \{1,2\}$, $\ket{1} = \fcre{a} \fani{i} \ket{0_\text{e} 0_\text{B}}$, $\ket{2} = \fcre{a} \fani{i} \bcre{\nu} \ket{0_\text{e} 0_\text{B}}$, $\bra{1} = \left( \ket{1} \right)^\dagger$, $\bra{2} = \left( \ket{2} \right)^\dagger$, and $\ket{0} \equiv \ket{0_\text{e} 0_\text{B}}$ denotes the reference ground-state wave function.

Within the perturbative expansion, the $n$th-order contribution to the one-body density matrix is obtained from 
\begin{equation}
	\gamma_{pq}^{(n)} = \pdv{E^{(n)}}{f_{pq}}
\end{equation}
Under certain approximations, the first-order density matrix contributions of the occupied-occupied and virtual-virtual blocks is obtained from the second-order energy expression 
\begin{equation}
	E^{(2)} 
	= \bra{0} \comm{ \hV }{ \hT^{(1)}} 
	+ \comm{ \hZ^{(1)} }{ \hV } 
	+ \comm{ \hZ^{(1)} }{ \comm{ \hat{H}_0 }{ \hT^{(1)} } } \ket{0}
\end{equation}
The resulting density matrix blocks coincide with the linearized one-body $GW$ density matrix [see Eqs.~\eqref{eq:OccOccBlockD} and \eqref{eq:VirtVirtBlockD}] only when canonical Hartree--Fock orbitals are employed.
If this is not the case, additional terms, due to non-zero amplitudes $t^{(1)}_{ia}$ and $z^{(1)}_{ia}$ have to be considered in these blocks as well.
Furthermore, only the first term of the occupied-virtual block [see Eq.~\eqref{eq:OccVirtBlockD}] is obtained at second-order, which is zero for canonical orbitals. 

The additional terms of Eq.~\eqref{eq:OccVirtBlockD} are recovered from a modified third-order energy expression
\begin{equation}
	E^{(3)} 
	= \bra{0} \comm{ \hV }{ \hT^{(2)}_1 }
	+ \comm{ \hZ^{(2)}_1 }{ \hat{V} } \ket{0} 
\end{equation}
Note that the resulting density matrix is not symmetric, i.e., $\gamma_{pq} \neq \gamma_{qp}$, and is symmetrized throughout this work for convenience.
The natural occupation numbers from the perturbative ECC treatment then coincide with those obtained from the linearized one-body $GW$ density matrix, as reported in Eqs.~\eqref{eq:OccOccBlockD}--\eqref{eq:OccVirtBlockD}.
Complete spin-adapted working equations are provided in the \SupInf.

\section{Computational Details}
\label{sec:compdet}

All calculations were performed using a custom implementation built on the \textsc{PySCF} package.\cite{Sun_2018,Sun_2020}
Unless otherwise stated, $G_0W_0$ calculations were carried out without invoking the diagonal approximation, and all results employ Hartree--Fock as the mean-field starting point.
The linearized $GW$ density matrix was computed using the perturbative ECC framework described in Sec.~\ref{sec:GWLDM}.
The test set consists of 23 small molecules with accurate theoretical best estimates (TBEs) for inner- and outer-valence ionization potentials (IPs), taken from Ref.~\onlinecite{Marie_2024b}.
All calculations employ the aug-cc-pVQZ basis set.\cite{Dunning_1989,Woon_1993}
We employ a large basis set to reduce basis set effects, thereby facilitating a more consistent, like-for-like comparison between the different schemes.
The self-consistent field procedure is converged until the total energy changes by less than \SI{d-9}{\hartree}. 
ECC amplitudes are iterated until the update satisfies $\norm{\Delta \bt} < \num{d-8}$ and $\norm{\Delta \bz} < \num{d-8}$.
The Davidson solver used for computing $G_0W_0$ quasiparticle energies is converged to \SI{d-8}{\hartree}.

\section{Numerical results}
\label{sec:results}

\begin{figure*}
	\includegraphics[width=1\textwidth]{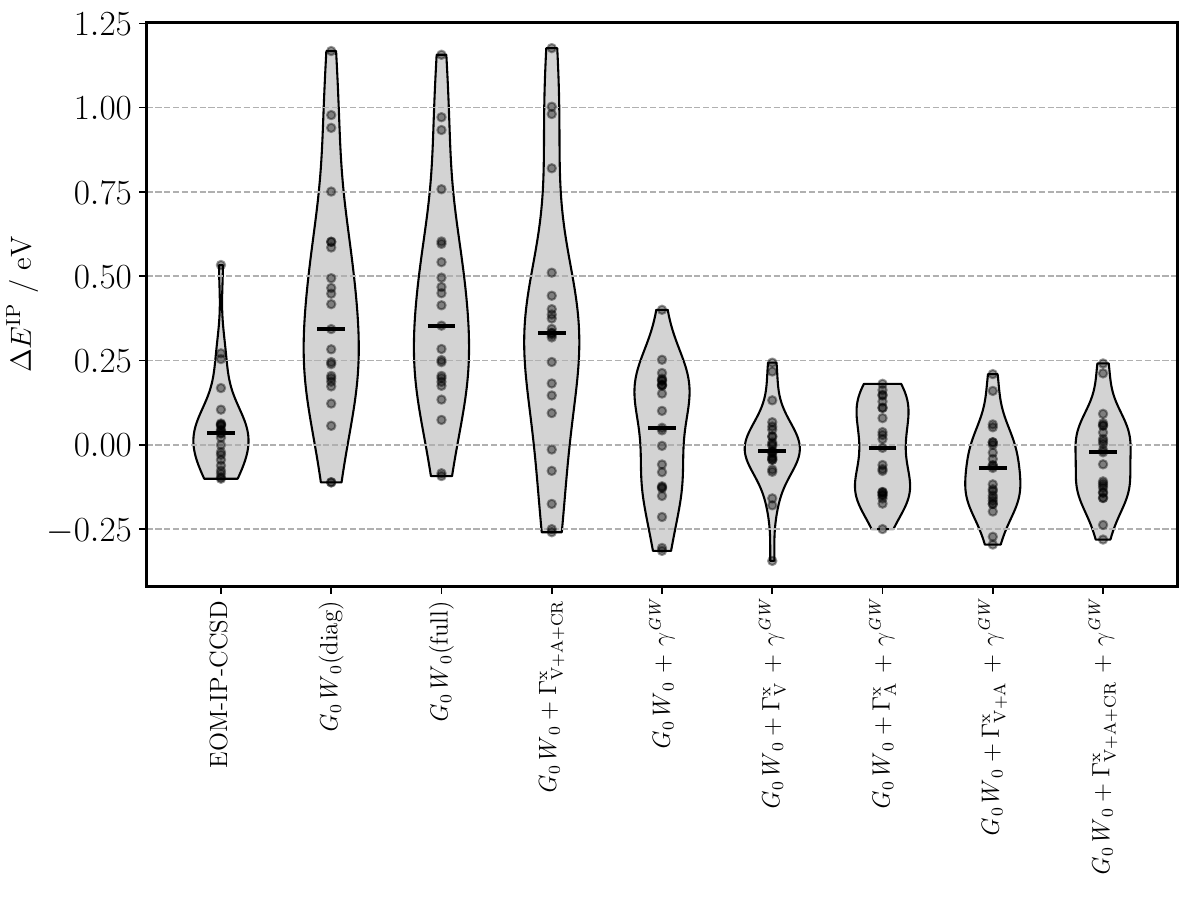} 
	\caption{Violin plots of the errors (in \si{\eV}) for the principal IPs obtained with various $G_0W_0$ variants and EOM-IP-CCSD with respect to the TBEs for the benchmark set of 23 molecular systems from Ref.~\citenum{Marie_2024b} computed with the aug-cc-pVQZ basis.
	Note that $\Gamma_\text{V+A}^\text{x} = \Gamma_\text{V}^\text{x} + \Gamma_\text{A}^\text{x}$ and $\Gamma_\text{V+A+CR}^\text{x} = \Gamma_\text{V+A}^\text{x} + \Gamma_\text{CR}^\text{x}$.}
	\label{fig:ips_val}
\end{figure*}

First, we investigate the performance of various $G_0W_0$ vertex corrections, as proposed in Sec.~\ref{sec:Vertexcorrections}, for the calculation of the principal IPs (i.e., the lowest-energy IP of each system).
The with mean absolute errors (MAEs) and mean-signed errors (MSEs) for these 23 principal IPs with respect to the reference TBE values for the benchmark set of Ref.~\citenum{Marie_2024b} are displayed in Table \ref{tab:ips_val}.
The relative errors are provided in the \SupInf, and the error distributions are visualized in the violin plots of Fig.~\ref{fig:ips_val}.

In total, eight distinct $G_0W_0$ variants are considered.
First, $G_0W_0$ IPs are computed within the diagonal approximation (as reported in Ref.~\citenum{Marie_2024b}), denoted as $G_0W_0$ (diag), as well as using the full self-energy, denoted as $G_0W_0$ (full). 
Furthermore, we consider $G_0W_0$ calculations including static Fock matrix corrections using the $GW$ linearized density matrix (denoted as $+\gamma^{GW}$).
Finally, we explore the effect of vertex corrections in $\bH^\EOM$: (i) $\Gamma_\text{V}^\text{x}$, (ii) $\Gamma_\text{A}^\text{x}$, and (iii) $\Gamma_\text{CR}^\text{x}$.
For a motivation of these contributions, the reader is referred to Sec.~\ref{sec:Vertexcorrections}.
Combinations of these vertex corrections are also considered and are denoted as, for example, $\Gamma^{\text{x}}_\text{V+A}$, corresponding to $\Gamma^{\text{x}}_\text{V}+ \Gamma^{\text{x}}_\text{A}$.
Vertex corrections are also considered in combination with static Fock matrix corrections.

Overall, similar errors in the principal IPs are observed for both $G_0W_0$ (diag) and $G_0W_0$ (full), MAEs of \SI{0.423}{\eV} and \SI{0.420}{\eV}, respectively.
In general, the IPs for $G_0W_0$ (diag) and $G_0W_0$ (full) in combination with the Hartree--Fock starting point are systematically overestimated when compared to the TBEs.
This is a known issue for $G_0W_0$, \cite{Stan_2006,Caruso_2012,Caruso_2013a,Caruso_2016,Maggio_2016,Wen_2024} and by performing full self-consistency, the IPs are known to be lowered. \cite{Stan_2006,Caruso_2012,Caruso_2013a,Caruso_2016,Wen_2024}
We observe a similar trend here, where the inclusion of static Fock matrix corrections based on the $GW$ linearized density matrix lowers the IPs, resulting in a MAE of \SI{0.166}{\eV}.
A similar trend has been observed in \revtwo{Ref.~\citenum{Bruneval_2021}.}

Including a combination of all three vertex corrections, $\Gamma_\text{V+A+CR}^\text{x}$, does only slightly improve the results, compared to $G_0W_0$, and yields a MAE of \SI{0.420}{\eV}.
However, when combined with static Fock matrix corrections, the MAE is significantly reduced. 
While all variants including vertex corrections in combination with static Fock matrix corrections ($+\Gamma_\text{V}^\text{x} + \gamma^{GW}$, $+\Gamma_\text{A}^\text{x} + \gamma^{GW}$, $+\Gamma_\text{V+A}^\text{x} + \gamma^{GW}$, $+\Gamma_\text{V+A+CR}^\text{x} + \gamma^{GW}$) result in improvements when compared to $G_0W_0 + \gamma^{GW}$, the best performance is obtained when $\Gamma_\text{V}^\text{x}$ is considered.
In this case one finds a MAE of \SI{0.082}{\eV}, and a MSE of \SI{-0.011}{\eV}, which is even better than the MAE of \SI{0.098}{\eV} and MSE of \SI{-0.053}{\eV} obtained from EOM-IP-CCSD.
All variants lower the magnitude of the IPs slightly further when compared to $G_0W_0 + \gamma^{GW}$, while also reducing the spread of the errors.
\rev{This reflects the limitations of HF orbitals as a starting point for the $G_0W_0$ calculation. 
We conjecture that similar improvements could be achieved using optimally tuned range-separated hybrid orbitals.\cite{McKeon_2022}}

A direct comparison between the present results and previously proposed vertex corrections for molecular IPs is challenging. 
The available studies differ in several important aspects, including the molecular test sets considered, \cite{Vlcek_2019,Wen_2024,Bruneval_2025} the underlying mean-field starting points, \cite{Ren_2015} as well as the basis sets and other numerical settings employed. \cite{Maggio_2017b,Vlcek_2019}
Even qualitative trends are difficult to establish: while some vertex-corrected schemes systematically decrease IPs, \cite{Maggio_2016,Vlcek_2019,Forster_2025} others lead to an overall increase. \cite{Ren_2015,Wen_2024} 
\rev{Recent evidence suggests that inclusion of vertex corrections only in $P$ tends to reduce the quasiparticle gap and lower the IPs, whereas the inclusion of vertex corrections only in $\Sigma$ tends to increase both the gap and the IPs. \cite{Cunningham_2018,Bruneval_2024,Forster_2024,Forster_2025,Cunningham_2023}
Achieving systematic improvements appears to require incorporating vertex corrections in both $P$ and $\Sigma$. \cite{Forster_2024,Forster_2025}}
Consequently, it is not straightforward to provide a definitive assessment of the present vertex corrections relative to alternative approaches.
Nevertheless, we emphasize that the vertex corrections introduced here lead to a clear and significant improvement in the description of principal IPs.

\begin{squeezetable}
\begin{table*}
\begin{ruledtabular}
\caption{MAE and MSE (in \si{\eV}) of the principal IPs with respect to the TBEs for the benchmark set of 23 molecular systems from Ref.~\onlinecite{Marie_2024b} computed using various $G_0W_0$ variants with the aug-cc-pVQZ basis.}
\label{tab:ips_val}
\begin{tabular}{l r r r r r r r r r}
	& & \multicolumn{8}{c}{$G_0W_0$ variants} \\
	\cline{3-10} \vspace{-0.2cm}\\ 
	 & CCSD\fnm[1] & 
	(diag)\fnm[2] &
	(full) &
	+ $\Gamma^{\text{x}}_\text{A+V+CR}$ &
	+ $\gamma^{GW}$ &
	$+ \Gamma^{\text{x}}_\text{V}$ + $\gamma^{GW}$ &
	$\Gamma^{\text{x}}_\text{A}$ + $\gamma^{GW}$ &
	$+ \Gamma^{\text{x}}_\text{V+A}$ + $\gamma^{GW}$ &
	$+\Gamma^{\text{x}}_\text{V+A+CR}$ + $\gamma^{GW}$ \vspace{0.1cm}\\
	\cline{1-10}\vspace{-0.2cm}\\
MAE & 0.098 & 0.423 & 0.420 & 0.400 & 0.16\revtwo{6} & 0.082 & 0.116 & 0.119 & 0.108 \\
MSE & 0.053 & 0.403 & 0.405 & 0.332 & 0.03\revtwo{5} & -0.011 & -0.017 & -0.076 & -0.039 \\
\end{tabular}
\end{ruledtabular}
\fnt[1]{EOM-IP-CCSD data taken from Ref.~\citenum{Marie_2024b}.}
\fnt[2]{$G_0W_0$ results computed within the diagonal approximation taken from Ref.~\citenum{Marie_2024b}.}
\end{table*}
\end{squeezetable}

\begin{figure*}
	\includegraphics[width=1\textwidth]{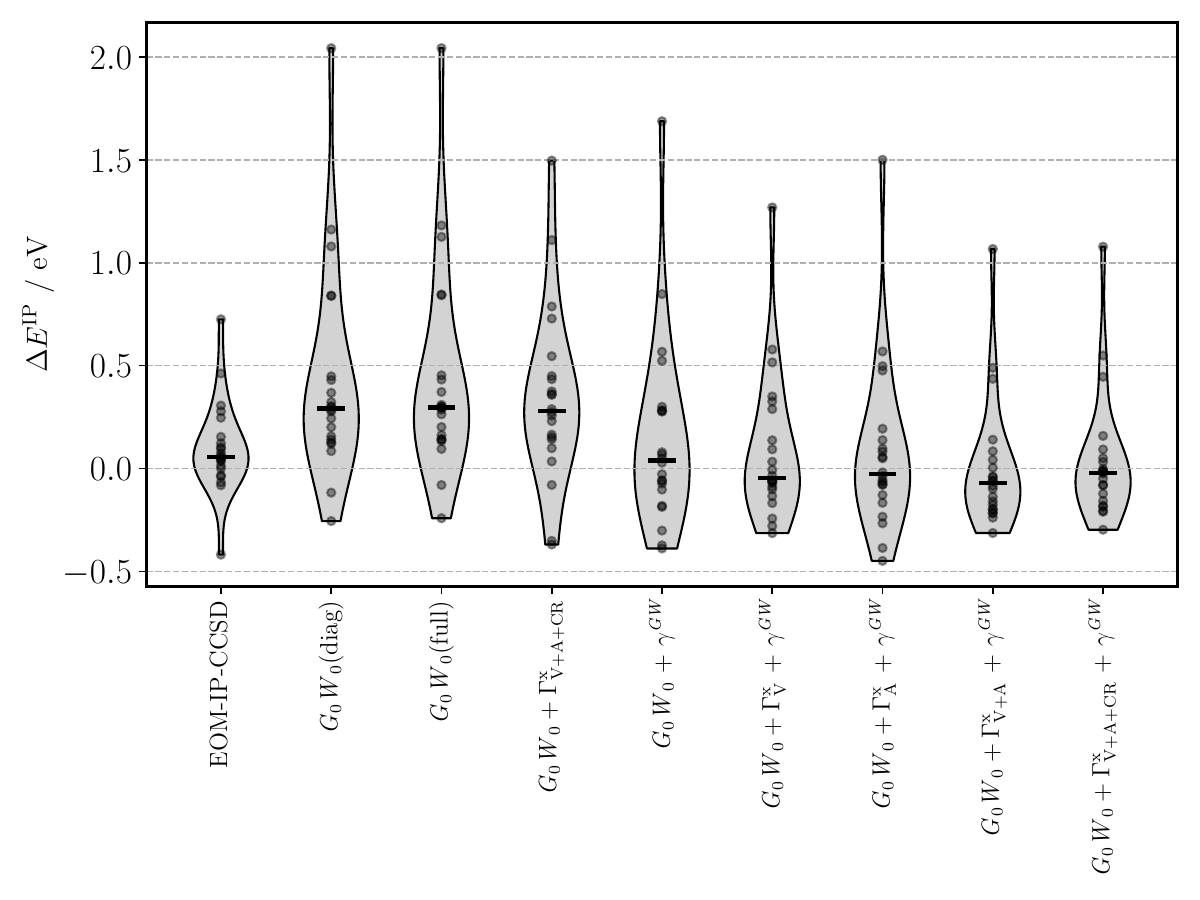} 
	\caption{Violin plots of the errors (in \si{\eV}) for the second IPs obtained with various $G_0W_0$ variants and EOM-IP-CCSD with respect to the TBEs for the benchmark set of 23 molecular systems from Ref.~\citenum{Marie_2024b} computed with the aug-cc-pVQZ basis. 	Note that $\Gamma_\text{V+A}^\text{x} = \Gamma_\text{V}^\text{x} + \Gamma_\text{A}^\text{x}$ and $\Gamma_\text{V+A+CR}^\text{x} = \Gamma_\text{V+A}^\text{x} + \Gamma_\text{CR}^\text{x}$.}
	\label{fig:ips_2}
\end{figure*}

Next, we assess the performance of the different $G_0W_0$ variants for the second IPs. 
The error distributions are visualized in the violin plots of Fig.~\ref{fig:ips_2}, and 
MAEs and MSEs are displayed in Table \ref{tab:ips_2}.
The relative errors are provided in the \SupInf.
For the second IPs, we observe overall similar trends as for the principal IPs.
While $G_0W_0$ (diag) and $G_0W_0$ (full) yield similar MAEs of \SI{0.474}{\eV} and \SI{0.466}{\eV}, the inclusion of static Fock matrix corrections through the $GW$ linearized density reduces the MAE to \SI{0.306}{\eV}.
Notably, all vertex corrections in combination with $\gamma^{GW}$ yield significant improvements of more than \SI{0.7}{\eV} when compared to $G_0W_0 + \gamma^{GW}$.
The best overall performance is achieved with $+\Gamma_\text{V+A+CR}^\text{x} + \gamma^{GW}$, resulting in a MAE of \SI{0.183}{\eV}, only slightly higher than the MAE of \SI{0.157}{\eV} obtained from EOM-IP-CCSD.
Moreover, the corresponding MSE of \SI{0.035}{\eV} is lower than the MSE of \SI{0.099}{\eV} obtained for EOM-IP-CCSD.
In all cases, the largest outlier is for Argon.
Given the large reference TBE value for the second IP (\SI{29.182}{\eV}), the relative error of \SI{2.48}{\percent} for EOM-IP-CCSD and \SI{3.69}{\percent} for $G_0W_0 +\Gamma_\text{V+A+CR}^\text{x} + \gamma^{GW}$ is relatively small compared to the absolute error of \SI{0.725}{\eV} and \SI{1.078}{\eV}, respectively.

\begin{squeezetable}
\begin{table*}
\begin{ruledtabular}
	\caption{MAE and MSE (in \si{\eV}) of the second IPs with respect to the TBEs for the benchmark set of 23 molecular systems from Ref.~\onlinecite{Marie_2024b} computed using various $G_0W_0$ variants with the aug-cc-pVQZ basis.}
	\label{tab:ips_2}
	\begin{tabular}{l r r r r r r r r r}
		& & \multicolumn{8}{c}{$G_0W_0$ variants} \\
		\cline{3-10} \vspace{-0.2cm}\\ 
		 & CCSD\fnm[1] & 
		(diag)\fnm[2] &
		(full) &
		+ $\Gamma^{\text{x}}_\text{A+V+CR}$ &
		+ $\gamma^{GW}$ &
		$+ \Gamma^{\text{x}}_\text{V}$ + $\gamma^{GW}$ &
		$\Gamma^{\text{x}}_\text{A}$ + $\gamma^{GW}$ &
		$+ \Gamma^{\text{x}}_\text{V+A}$ + $\gamma^{GW}$ &
		$+\Gamma^{\text{x}}_\text{V+A+CR}$ + $\gamma^{GW}$ \vspace{0.1cm}\\
		\cline{1-10}\vspace{-0.2cm}\\
		\hline \vspace{-0.2cm}\\
		MAE & 0.157 & 0.461 & 0.466 & 0.413 & 0.30\revtwo{6} & 0.237 & 0.255 & 0.204 & 0.184 \\
		MSE & 0.099 & 0.427 & 0.437 & 0.341 & 0.14\revtwo{7} & 0.090 & 0.077 & 0.001 & 0.035 \\
	\end{tabular}
	\end{ruledtabular}
	\fnt[1]{EOM-IP-CCSD data taken from Ref.~\citenum{Marie_2024b}.}
	\fnt[2]{$G_0W_0$ results computed within the diagonal approximation taken from Ref.~\citenum{Marie_2024b}.}
\end{table*}
\end{squeezetable}

\section{Conclusion}
\label{sec:conclusion}
In summary, we have established a formal connection between the extended direct ring coupled cluster doubles (ECCD) framework and the $G_0W_0$ approximation.
When applied to the electron-boson Hamiltonian in combination with the equation-of-motion formalism, this procedure recovers the $G_0W_0$ quasiparticle energies exactly.
Furthermore, we have derived the linearized one-body $GW$ density matrix within the ECC framework using perturbation theory.
Exploiting the established connection between ECC and $GW$, we have proposed several vertex corrections beyond $GW$ within the ECC framework,
while \rev{retaining a sum-over-state presentation of the corresponding self-energy.} 
Preliminary numerical results for a benchmark set of 23 small molecules demonstrate the potential of the proposed vertex corrections to significantly improve the accuracy of $G_0W_0$ ionization potentials.
Overall, the present work opens exciting avenues for including vertex corrections beyond $GW$ within the ECC framework, such as the inclusion
of particle-particle and hole-hole ladder contributions.

\section*{Supporting Information}
The \SupInf contains the derivations and working equations supporting the connection between the $GW$ approximation and the extended coupled-cluster framework (key amplitude identities, the full spin-adapted EOM $\sigma$-vector equations, including the $G_0W_0$–TDA subset and ECC-based vertex corrections, and the perturbative ECC formulation of the linearized one-body $GW$ density matrix). 
It also reports the errors of the principal IPs with respect to the TBEs for the benchmark set of 23 molecular systems.

\section*{Acknowledgements}
The authors would like to thank Antoine Marie for insightful discussions.
This project has received funding from the European Research Council (ERC) under the European Union's Horizon 2020 research and innovation programme (Grant agreement No.~863481).
J.~T.~acknowledges funding from the Fonds der Chemischen Industrie (FCI) via a Liebig fellowship and support by the Cluster of Excellence ``CUI: Advanced Imaging of Matter'' of the Deutsche Forschungsgemeinschaft (DFG) (EXC 2056, funding ID 390715994).
For this work, the HPC-cluster Hummel-2 at the University of Hamburg was used. 
The cluster was funded by Deutsche Forschungsgemeinschaft (DFG, German Research Foundation) - 498394658.
Support and funding from the European Research Council (ERC) (Grant Agreement No. 101087184) is gratefully acknowledged.


\section*{References}

\end{document}